\documentclass[11t, a4paper]{article}
\pdfoutput=1
\usepackage{jcappub}
\usepackage[utf8]{inputenc}
\usepackage[english]{babel}
\usepackage{amsmath, amssymb}
\usepackage{mathtools}
\usepackage{graphicx}
\usepackage{hyperref}

\newcommand*{\e}{\mathrm{e}}
\newcommand*{\im}{\mathrm{i}}

\makeatletter
\newcommand*{\dd}{\mathop{}\!{\operator@font d}}
\makeatother

\newcommand*{\mH}{\mathcal{H}}

\newcommand*{\vk}{\mathbf{k}}
\newcommand*{\vq}{\mathbf{q}}
\newcommand*{\vp}{\mathbf{p}}
\newcommand*{\vs}{\mathbf{s}}
\newcommand*{\vu}{\mathbf{u}}
\newcommand*{\vv}{\mathbf{v}}
\newcommand*{\vx}{\mathbf{x}}

\newcommand*{\zhat}{\hat{z}}

\newcommand*{\ihMpc}{h\ \text{Mpc}^{-1}}
\newcommand*{\Mpc}{\mathrm{Mpc}}

\newcommand*{\Pnw}{P_{\mathrm{nw}}}
\newcommand*{\Pw}{P_{\mathrm{w}}}

\newcommand*{\kmax}{k_{\mathrm{max}}}

\newcommand*{\bea}{\begin{eqnarray}}
\newcommand*{\eea}{\end{eqnarray}}
\newcommand*{\be}{\begin{equation}}
\newcommand*{\ee}{\end{equation}}
\newcommand*{\nn}{\nonumber}

\subheader{\small%\sc
    \vspace*{-1.28cm}
    \begin{flushright}
        TUM-HEP-1472/23
    \end{flushright}
}

\title{The two-loop power spectrum in redshift space}

\author[a]{Petter Taule,}

\affiliation[a]{Institut de physique th\'eorique, Universit\'e Paris Saclay CEA, CNRS, 91191 Gif-sur-Yvette, France}

\emailAdd{petter.taule@ipht.fr}

\author[b]{Mathias Garny}

\affiliation[b]{Physik Department T31, Technische Universit\"at M\"unchen, James-Franck-Stra\ss{}e 1, D-85748 Garching, Germany
}%

\emailAdd{mathias.garny@tum.de}

\abstract{%
    We present the matter power spectrum in redshift space including two-loop corrections. We follow a strictly perturbative approach incorporating all non-linearities
    entering both via the redshift-space mapping and within real space up to the required (fifth) order, complemented by suitable
    effective field theory (EFT) corrections. This approach can a priori be viable up to scales of order $0.2\ihMpc$ beyond which
    power suppression related to the finger-of-God effect becomes non-perturbatively strong.
    We extend a simplified treatment of EFT corrections at two-loop order from real to redshift space, making sure
    that the leading UV-sensitivity of both the single-hard and double-hard limit of the two-loop contributions to the power spectrum is accounted for, and featuring two
    free parameters for each multipole. Taking also infrared-resummation into account, we calibrate with and compare to Quijote $N$-body simulations for the monopole and quadrupole
    at redshifts $z=0$ and $z=0.5$. We find agreement within sample variance (at percent-level) up to $0.18\ihMpc$ at two-loop order, compared to $0.1\ihMpc$ at one-loop. We also
    investigate the role of higher-derivative corrections.
}

\begin{document}
\maketitle

\section{Introduction}

Galaxy surveys mapping out the three-dimensional large-scale structure (LSS)
of the Universe rely on redshift information to infer radial distances.
Hence theoretical interpretation of galaxy
clustering data is performed in \emph{redshift space} rather than real space,
accounting for the impact of galaxies' peculiar velocities along the line-of-sight, referred to as redshift-space distortions
(RSD)~\cite{Kaiser:1987qv}.  This greatly complicates the
theoretical modeling, but on the flip side RSD contain additional information
about peculiar velocities on cosmological scales, being sensitive to the linear growth rate $f = \dd \ln D/\dd \ln a$, where $D(z)$
is the linear growth factor (e.g.~\cite{BOSS:2016wmc}). Current galaxy surveys
such as DESI~\cite{DESI:2016fyo} and Euclid~\cite{Amendola:2016saw} will
measure RSD at unprecedented precision, yielding percent-level measurements of
the growth rate~\cite{Amendola:2016saw}.

On mildly non-linear scales the clustering of matter in
real space can be described perturbatively within a fluid
model~\cite{Bernardeau:2001qr}, following the evolution of the lowest moments
of the Boltzmann hierarchy: the density and velocity fields. Given its initial
smallness, the second moment --- the velocity dispersion tensor --- can be neglected on very
large scales. It is however well-known that non-linear evolution leads to
sizable velocity dispersion, which back-reacts on mildly non-linear scales via
mode-coupling~\cite{Pueblas:2008uv}. Hence an accurate prescription for the
dispersion tensor is required. In the effective field theory (EFT) approach, the velocity dispersion tensor is
treated as a functional of the long-wavelength fields (and gradients of them), including all combinations
 that are allowed by symmetries, encompassing mass and momentum conservation as well as extended Galilean invariance~\cite{Baumann:2010tm}. Each contribution can be written as a product of an operator
containing the fields and an unknown Wilson coefficient (frequently dubbed counterterm in this context).
The latter is not predicted by the theory but rather inferred by fitting to $N$-body simulations or marginalized over in data
analysis. Although introducing additional free coefficients, the EFT approach has
proven a useful alternative to exploit the full-shape information~\cite{Gil-Marin:2015sqa,BOSS:2016psr,BOSS:2016teh,BOSS:2016ntk,BOSS:2016off,Troster:2019ean,Semenaite:2021aen}
of BOSS clustering data~\cite{BOSS:2016wmc}, see~\cite{Ivanov:2019pdj, DAmico:2019fhj, Chen:2021wdi}. Apart from an EFT description, the velocity dispersion tensor can also be included as a
dynamic variable, trading the need for free coefficients for a greater computational complexity~\cite{Garny:2022tlk,
Garny:2022kbk}.

Mapping the density field to redshift space has two main
effects~\cite{Kaiser:1987qv, Jackson:1971sky, Scoccimarro:1999ed, Scoccimarro:2004tg}: large-scale
overdensities will look squashed along the line-of-sight due to the infalling
peculiar velocity field, yielding enhanced power on large separations
parallel to the line-of-sight. On the other hand, large pairwise velocities on
small scales lead to distortions being elongated parallel to the
line-of-sight, referred to as the ``fingers-of-God'' (FoG) effect~\cite{Jackson:1971sky}. The former can be
adequately described in linear theory with the well-known Kaiser
formula~\cite{Kaiser:1987qv} and treated perturbatively at higher order. Due to its non-linear nature, the second effect
presents a much greater challenge for a perturbative treatment.
In particular, the distribution of pairwise velocities determining the suppression of the redshift-space power spectrum on the FoG scale $k\sim 0.2\ihMpc$ features
strongly non-Gaussian tails generated by non-linear evolution~\cite{Scoccimarro:2004tg}. This motivates an ansatz for the suppression based on a combination
of Gaussian and Lorentzian form~\cite{BOSS:2016off, Eggemeier:2022anw}, encapsulating the impact of higher moments of the pairwise velocity distribution.
Following instead a strictly perturbative approach, as done in this work, the mapping to redshift space involves an additional power series
expansion of velocity fields, increasing the relative importance of higher loop corrections
on mildly non-linear scales as compared to the power spectrum in real space. In addition, the mapping involves products of
fields at the same spatial position, giving rise to contact terms, being related to the pairwise velocity moments at
infinite separation~\cite{Scoccimarro:2004tg}. Within the EFT approach these need to
be renormalized by additional counterterms~\cite{Desjacques:2018pfv,Senatore:2014vja} (see also~\cite{Chen:2020fxs}).
A priori a strictly perturbative approach in redshift space can be expected to be in principle possible for scales up to around the FoG scale $k\lesssim 0.2\ihMpc$.

\begin{figure}[t]
    \centering
    \includegraphics[width=0.8\textwidth]{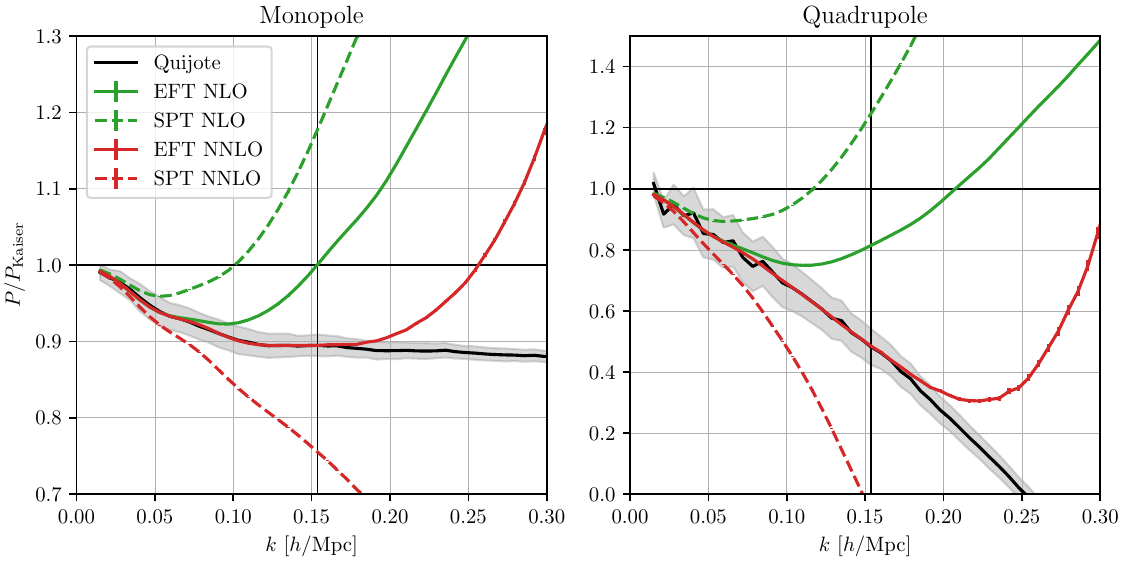}
    \caption{Perturbation theory predictions for the matter power spectrum monopole
        and quadrupole in redshift space, compared to Quijote $N$-body simulation
        data at redshift $z = 0$ (black). The graphs are normalized to the linear
        Kaiser prediction (see Eq.~\eqref{eq:integrand}). Gray bands indicate the $N$-body
        simulation uncertainty from sample variance. The green solid lines correspond to a 0-parameter
        NLO prediction (using the ``$c_s^2$'' EFT coefficient measured at
        two-loop, rather than fitting it at one-loop, in order to mitigate overfitting issues at one-loop, see
        Sec.~\ref{sec:numerics}). The red solid lines show our 2-parameter NNLO results.
        The corresponding SPT results (no EFT corrections) are displayed with dashed lines.
        The vertical line indicates the pivot scale $k_\text{max}$.}
    \label{fig:pk_vs_kaiser}
\end{figure}

In this work, we present the complete \emph{two-loop} matter power
spectrum in redshift space in the EFT approach, for the first time to our knowledge%
\footnote{The two-loop redshift-space power spectrum is calulated within the
    \emph{GridSPT} framework in~\cite{Taruya:2021ftd}.
    See also~\cite{Taruya:2013my}, where two-loop contributions in redshift space are discussed
in the context of the TNS model~\cite{Taruya:2010mx}. We note that the latter
approach does not include all terms contributing in a strict perturbative
expansion of the redshift-space mapping, but instead encapsulates some of them
into a finger-of-God factor that is not treated perturbatively, but replaced by
a phenomenological suppression function.}.
We follow a perturbative approach complemented by EFT corrections, involving the density field in redshift space up to
fifth order. We compute the ultraviolet (UV) sensitivity of the redshift-space two-loop
correction analytically for EdS kernels, establishing that the
double-hard limit can be renormalized by the counterterms present at one-loop
(as required by mass and momentum conservation). We add one extra EFT parameter
for each redshift-space multipole to renormalize the single-hard limit of the
two-loop correction, making the ansatz that the small-scale physics affect
intermediate scales in a manner matching the UV-sensitivity of the bare
theory, which has been shown to work well for the two-loop power spectrum in real space~\cite{Baldauf:2015aha}.
Thus, at next-to-next-to leading order (NNLO), we have two free
coefficients for each multipole. After implementing infrared (IR) resummation, we
calibrate and compare our predictions to Quijote $N$-body simulation data.

The main result of this work is displayed in Fig.~\ref{fig:pk_vs_kaiser}. The
perturbation theory predictions for the redshift-space monopole and quadrupole
at redshift $z = 0$ at NLO and NNLO are shown, as well as the Quijote $N$-body
simulation result. Here, we used a 0-parameter model at NLO and a 2-parameter model
(for each multipole) at NNLO (for the NLO counterterms, we use the value of
the corresponding EFT parameters more precisely measured at NNLO, see
Sec.~\ref{sec:numerics}).
For comparison, the SPT results are displayed with dashed lines.
We find that the NLO EFT result is in agreement with the Quijote $N$-body data (within uncertainties from sample variance)
up to $k \simeq 0.1\ihMpc$, while adding the two-loop increases the wavenumbers with agreement to $k \simeq 0.18\ihMpc$ at NNLO.

This work is structured as follows: in Sec.~\ref{sec:pt}, we review Eulerian
perturbation theory in redshift space, and outline the EFT framework we
utilize. Moreover, we demonstrate the IR-resummation procedure we adopt. In
Sec.~\ref{sec:numerics}, we fit and compare our perturbative results to Quijote $N$-body
data, contrasting the NNLO result with various prescriptions for treating the NLO case. We conclude in
Sec.~\ref{sec:conclusion}. In App.~\ref{app:kernels} we describe our implementation of $Z_n$ kernels for the matter density contrast in redshift space.
We perform various tests to validate our method as well as provide some supplementary results in App.~\ref{app:validation}.
Finally, in App.~\ref{app:singledoublehard} we present analytical expressions for the single- and double-hard limit of the
two-loop power spectrum monopole and quadrupole in redshift space, as well as for the corresponding two-loop EFT corrections.

\section{Perturbation theory in redshift space}
\label{sec:pt}

In this section, we provide a brief review of Eulerian perturbation theory in
redshift space, establishing the formalism and notation we use. Furthermore, we
describe the EFT framework that we will utilize at two-loop order.

The distance to observed galaxies is inferred from their radial velocities,
which includes contributions from both the Hubble flow, reflecting the true
distance, as well as ``distortions'' caused by the galaxies' peculiar
velocities~\cite{Kaiser:1987qv, Bernardeau:2001qr, Scoccimarro:2004tg}. An object at position $\vx$
is assigned the \emph{redshift space} coordinate
\begin{equation}
    \vs=\vx - f u_z(\vx) \zhat,
    \label{eq:rsd_def}
\end{equation}
where $f = \dd \ln D/\dd \ln a$ is the growth rate,
$\vu(\vx) = - \vv(\vx)/\mH f$ denotes the (rescaled) peculiar velocity, and we
have assumed that the line-of-sight is a fixed direction $\zhat$
(plane-parallel approximation). By using the preservation of mass density under
the coordinate transformation, one can show that the redshift space density
contrast $\delta_s$ in Fourier space is given by~\cite{Bernardeau:2001qr}%
\footnote{Neglecting contributions for which $f \nabla_z u_z(\vx) \geq 1$.}
\begin{equation}
    \delta_s(\vk) =
    \int \frac{\dd^3\vx}{(2\pi)^3}
    \e^{-\im \vk\cdot\vx}
    \e^{\im f k_z u_z(\vx)}
    \left[
        \delta(\vx) + f \nabla_z u_z(\vx)
    \right]
    \,.
    \label{eq:delta_s}
\end{equation}
Expanding the exponential function as well as the fields themselves yields the
following perturbative expression
\begin{align}
    \delta_s(\vk) &=
    \sum_{n=1}^{\infty} \int_{\vq_1,\dotsc,\vq_n}
    \delta_D
    \left(
        \vk - \sum_{i=1}^{n} \vq_{i}
    \right)
    \left[
        \delta(\vq_1) + f \mu_1^2 \theta(\vq_1)
    \right]
    \frac{ (f \mu k)^{n-1}}{(n-1)!}
    \,
    \frac{\mu_2}{q_2}\theta(\vq_2) \dotsb
    \frac{\mu_n}{q_n}\theta(\vq_n)
    \nonumber \\
    & \equiv
    \sum_{n=1}^{\infty}
    \int_{\vq_1,\dotsc,\vq_n}
    \delta_D
    \left(
        \vk - \sum_{i=1}^{n} \vq_{i}
    \right)
    \,
    Z_n (\vq_1, \dotsc, \vq_n) \,
    \delta_1(\vq_1) \dotsb \delta_1(\vq_n)
    \,,
    \label{eq:Z_n}
\end{align}
where we adopted the notation $\int_{\vq} = \int \dd^3 \vq$.
The redshift space kernels $Z_n$ represent the perturbative expansion, and they
are given up to third order as
\begin{subequations}
\begin{align}
    Z_1 (\vq) &= (1+f \mu^2)
    \,, \\
    Z_2 (\vq_1,\vq_2) &= F_2 (\vq_1,\vq_2) +f \mu^2 G_2 (\vq_1,\vq_2)
        + \frac{f \mu k}{2}
        \left[
            \frac{\mu_1}{q_1} (1+f \mu_2 ^2)
            + \frac{\mu_2}{q_2}(1+f \mu_1 ^2)
        \right]
    \,, \\
    Z_3(\vq_1,\vq_2,\vq_3) &=
        F_3 (\vq_1,\vq_2,\vq_3) +f \mu^2 G_3 (\vq_1,\vq_2,\vq_3)
        \nonumber \\
        & \phantom{=}
         + f \mu k
         \left[
             F_2 (\vq_1,\vq_2) +f \mu_{12}^2 G_2 (\vq_1,\vq_2)
         \right]
        \frac{\mu_3}{q_3}
        + f \mu k (1+f \mu_1 ^2) \frac{\mu_{23}}{q_{23}} G_2(\vq_2,\vq_3)
        \nonumber \\
        & \phantom{=}
        + \frac{(f \mu k)^2}{2} (1+f \mu_1 ^2) \frac{\mu_2}{q_2} \frac{\mu_3}{q_3}
        + \text{perm.}\,.
\end{align}
\end{subequations}
In the above expressions,
$\vk = \sum_i \vq_i$, $\mu = \vk \cdot \zhat/k$, $\mu_i = \vq_i \cdot \zhat/q_i$
and $F_n$ and $G_n$ denote the usual (symmetrized)
density contrast and velocity divergence kernels, respectively. For brevity, we have
omitted the explicit contributions from two permutations of wavenumber
arguments in the second and third line of the expression for $Z_3$. Note that the linear kernel is exactly
the Kaiser prediction.
In this work we use kernels up to $Z_5$. A general formula for the $n$-th order redshift-space kernel is given in App.~\ref{app:kernels}.
We note that the well-known constraints
from mass and momentum conservation as well as Galilean invariance translate into the properties
$Z_n(\vq_1,\dots,\vq_{n-2},\vq,-\vq)\propto k^2/q^2$ for $q\gg k$ and
\be\label{eq:Zsoft}
  Z_{n+1}(\vq_1,\dots,\vq_n,\vq)\to \frac{\vk\cdot\vq+fkq\mu\mu_q}{(n+1)q^2}Z_n(\vq_1,\dots,\vq_n)\,,
\ee
in the limit $q\to 0$, respectively. Here $\mu_q=(\vq\cdot\hat z)/q$. The former relation
is instrumental for the construction of the EFT corrections, and implies that they are largely
analogous as in real space, see Sec.~\ref{sec:eft}. The latter property ensures cancellation of contributions to loop integrals that are enhanced for small loop wavenumber, which
can be made explicit by the same algorithm as described in~\cite{Blas:2013bpa,Blas:2013aba}. It also underlies the
IR-resummation in redshift space~\cite{Ivanov:2018gjr}, see Sec.~\ref{sec:IRresum}.
In this work, our primary focus is
on the perturbative treatment of the mapping from real to redshift space,
therefore we do not consider bias, i.e.\ we model the total matter density in
redshift space.

Ultimately, we are interested in loop corrections to the power spectrum
multipoles in redshift space. To illustrate how we can numerically evaluate the
loop integrals as well as how we subsequently renormalize them, it is useful to
define the $L$-loop \emph{integrand} $p_{L}$ in redshift space as
\begin{equation}
    P_s^{L-\text{loop}}(k,\mu) =
    \int_{\vq_1,\dotsc,\vq_L}
    p_{s,L}(\vk,\mu,\vq_1,\dotsc,\vq_L)P_0(q_1)\cdots P_0(q_L)
    \,,
    \label{eq:P_L}
\end{equation}
where
\bea\label{eq:integrand}
  p_{s,L=0}(\vk,\mu) &=&  Z_1(\vk)^2P_0(k)\,,\nn\\
  p_{s,L=1}(\vk,\mu,\vq_1) &=&  6Z_3(\vk,\vq_1,-\vq_1)Z_1(\vk)P_0(k)+2Z_2(\vk-\vq_1,\vq_1)^2P_0(|\vk-\vq_1|)\,,\nn\\
  p_{s,L=2}(\vk,\mu,\vq_1,\vq_2) &=&  6Z_3(\vk-\vq_1-\vq_2,\vq_1,\vq_2)^2P_0(|\vk-\vq_1-\vq_2|)\nn\\
  && {} + \left(30Z_5(\vk,\vq_1,-\vq_1,\vq_2,-\vq_2)Z_1(\vk)+9Z_3(\vk,\vq_1,-\vq_1)Z_3(\vk,\vq_2,-\vq_2)\right)P_0(k)\nn\\
  && {} + 12Z_4(\vk-\vq_1,\vq_1,\vq_2,-\vq_2)Z_2(\vk-\vq_1,\vq_1)P_0(|\vk-\vq_1|)\nn\\
  && {} + 12Z_4(\vk-\vq_2,\vq_2,\vq_1,-\vq_1)Z_2(\vk-\vq_2,\vq_2)P_0(|\vk-\vq_2|)\,.
\eea
We use the freedom to relabel the integration variables to ensure $p_{s,2}(\vk,\mu,\vq_1,\vq_2)=p_{s,2}(\vk,\mu,\vq_2,\vq_1)$,
which is irrelevant for the two-loop result, but important for the definition of the two-loop EFT correction discussed below.
For numerical evaluation we in addition apply the rewritings as in~\cite{Blas:2013bpa,Blas:2013aba} to ensure cancellation
of IR enhanced terms at the integrand level as mentioned above.
The case $L=0$ corresponds to the linear Kaiser spectrum, $0-$loop$\equiv$lin, with no loop integration.

The power spectrum multipoles are projections onto Legendre polynomials
$\mathcal{P}_l(\mu)$:
\begin{equation}
    P_{l}^{L-\text{loop}}(k) =
    \frac{2l + 1}{2}
    \int_{-1}^{1} \dd\mu \, \mathcal{P}_l(\mu)
    \, P_s^{L-\text{loop}}(k,\mu)
    \,.
    \label{eq:P_l_L}
\end{equation}
We perform the loop integrals as well as the $\mu$-integral above numerically
using Monte Carlo integration. Hence, for each sample of the integrand, we
evaluate the kernels that appear at that order, which at two-loop includes
kernels up to $Z_5$. To compute the kernels efficiently, we develop an
extension of the algorithm for the $F_n$ and $G_n$ kernels from
Ref.~\cite{Blas:2013aba}, that in a recursive manner perturbatively expands the
various factors of the first line of Eq.~\eqref{eq:Z_n}. The algorithm is
explained in more detail in App.~\ref{app:kernels}.

\subsection{EFT framework}
\label{sec:eft}

It is well-known that SPT breaks down on even mildly non-linear scales and the
higher loop corrections suffer from large, spurious contributions from the
UV~\cite{Crocce:2005xy,Blas:2013aba}. To address this issue, within the EFT approach
``counterterms'' are incorporated that systematically
capture the impact of small scales on intermediate scales, renormalizing the
theory and restoring convergence of the perturbation
theory~\cite{Baumann:2010tm, Carrasco:2012cv}. In principle
all counterterms allowed by the symmetries of the problem (statistical homogeneity and
isotropy, mass and momentum conservation, extended Galilean invariance) should be included, accompanied by free
coefficients that are not predicted by the EFT. Moreover, the
mapping from real space to redshift space involves products of coarse-grained
fields at the same point in space (contact terms)~\cite{Scoccimarro:2004tg}, which in the EFT need to be
renormalized by additional counterterms~\cite{Desjacques:2018pfv,Senatore:2014vja}.

In practice, there are substantial degeneracies between
counterterms and one can only fit certain combinations of them accurately.
Furthermore, including many parameters increases the risk of overfitting when
comparing the EFT to data/simulations. For the two-loop power spectrum (in real
space), a two-parameter ansatz for the EFT corrections has been shown to work
well in practice~\cite{Baldauf:2015aha, Garny:2022fsh}. We briefly discuss this
scheme in the following and refer the reader to the quoted references for a
more thorough discussion. The renormalized two-loop power spectrum in real space
reads
\begin{equation}
    P_{\text{NNLO}}^{\text{ren.}}(k) =
    P^{\text{lin}}(k) +
    P^{\text{1-loop}}(k;\Lambda) +
    \bar{P}^{\text{2-loop}}(k;\Lambda) +
    \gamma_1(\Lambda) k^2 P^{\text{lin}}(k) +
    \gamma_2(\Lambda) \bar{P}_{\text{sh}}^{\text{2-loop}}(k;\Lambda)
    \,,
    \label{eq:P_2L_ren}
\end{equation}
where $P^{\text{lin}}\equiv P_0$ is the linear power spectrum and
$P^{L\text{-loop}}(\Lambda)$ is the (bare) $L$-th loop correction evaluated
with cutoff $\Lambda$. The bar notation will be explained shortly. The EFT
parameters are denoted by $\gamma_1$ and $\gamma_2$, with the former
representing the usual ``$c_s^2$'' correction to the sound speed of the fluid present
already at one-loop. The latter multiplies the \emph{single-hard} two-loop
contribution, defined implicitly as follows:
\bea
    P^{\text{2-loop}}(k;\Lambda)
    &\xrightarrow[q_1\gg k]{}&
    3 \int_{\vq_2}^\Lambda
    \int \frac{\dd \Omega_{q_1}}{4\pi}
    \left[
        q_1^2 \lim_{q_1\to\infty} p_{2}(\vk,\vq_1,\vq_2)P_0(q_2)
    \right]
    \frac{4\pi}{3} \int^{\Lambda} \dd q_1 P_0(q_1)\nn\\
    && {} \equiv
    \frac{1}{2} P_{\text{sh}}^{\text{2-loop}}(k;\Lambda) \sigma_d^2(\Lambda)
    \,,
    \label{eq:P_2L_sh}
\eea
where $p_2$ is the two-loop \emph{integrand} in real space and $\dd
\Omega_{q1}$ is the differential solid angle of $\vq_1$ (see App.~\ref{app:singledoublehard}). The above
factorization is valid at leading order in $k^2/q_1^2$ in the limit where $q_1$
is hard; due to the $1/q_1^2$-scaling of the kernels in the limit guaranteed by
momentum conservation, the bracket above is \emph{independent} of $\vq_1$, and
we could factorize out the displacement dispersion
$\sigma_d^2 \equiv 4\pi/3 \int \dd q\, P_0(q)$. The factor $1/2$ in the final
expression is present so that the definition of the single-hard contribution
$P_{\text{sh}}^{\text{2-loop}}$ also incorporates the equivalent limit in which
$q_2 \gg k, q_1$. Thus, this ansatz for the two-loop counterterm amounts to
assuming that the EFT corrects the UV in a universal manner dictated by the
UV-sensitivity of SPT. In other words, we assume that we can renormalize the
two-loop correction by shifting the value of the displacement dispersion
$\sigma_d^2 \to \sigma_d^2 + \gamma_2$.

Finally, the double-hard limit $q_1,q_2 \gg k$ of the two-loop correction is
renormalized by the $\gamma_1$-term introduced already at one-loop. We opt for
a renormalization scheme where $\gamma_1$ incorporates the correction of the
hard limit of the one-loop as well as the actual impact of short-scale physics
(the ``finite'' contribution), using the freedom to shift $\gamma_1$ such that
the double-hard contribution of the two-loop correction exactly cancels. In
practice, this means that we subtract the double hard contributions from the
two-loop correction $P^{\text{2-loop}}$ as well as from the single-hard
two-loop counterterm $P_{\text{sh}}^{\text{2-loop}}$. This operation is
indicated by bars in Eq.~\eqref{eq:P_2L_ren}. See
App.~\ref{app:singledoublehard} for analytic expressions for the single-
and double-hard limits of the two-loop.
We stress that this choice of subtracting the double-hard contribution has no
impact on the results, except for a shift in $\gamma_1$.

Next, we extend this EFT setup to redshift space. We renormalize each multipole
$l$ by the extension of the two-parameter ansatz described above,
\begin{equation}
    P_{l,\text{NNLO}}^{\text{ren.}}(k) =
    P_{l}^{\text{lin}}(k) +
    P_{l}^{\text{1-loop}}(k;\Lambda) +
    \bar{P}_{l}^{\text{2-loop}}(k;\Lambda) +
    \gamma_1(\Lambda) k^2 P_{l}^{\text{lin}}(k) +
    \gamma_2(\Lambda) \bar{P}_{l,\text{sh}}^{\text{2-loop}}(k;\Lambda)
    \,,
    \label{eq:P_l_2L_ren}
\end{equation}
where $P_{l}^{\text{lin}}$ is the Kaiser prediction and the loop corrections
are given by Eqs.~\eqref{eq:integrand} and \eqref{eq:P_l_L}. We define the single-hard two-loop
counterterm analogously to Eq.~\eqref{eq:P_2L_sh},
\begin{equation}
    P_{l,\text{sh}}^{\text{2-loop}}(k;\Lambda) =
    2\times 3 \times \frac{2l + 1}{2}
    \int_{-1}^{1} \dd\mu \, \mathcal{P}_l(\mu)
    \int_{\vq_2}
    \int \frac{\dd \Omega_{q_1}}{4\pi}
    \left[
        q_1^2 \lim_{q_1\to\infty} p_{s,2}(\vk,\mu,\vq_1,\vq_2)P_0(q_2)
    \right]
    \,,
    \label{eq:P_l_2L_sh}
\end{equation}
including a factor $2$ to account for the equivalent limit in which $\vq_2$ is
hard. Hence, we introduce two EFT parameters for each multipole in redshift
space. We give explicit expressions for the single-hard EFT correction $\bar{P}_{l,\text{sh}}^{\text{2-loop}}(k;\Lambda)$ in
App.~\ref{app:singledoublehard}, that effectively amount to one-loop integrals.
Furthermore, we demonstrate in App.~\ref{app:cutoff} that this prescription
yields cutoff-independent predictions for the multipoles. In practice, we find it convenient to compute the single-hard
limit numerically by fixing one wavenumber to a large value, $q_1 = 10\ihMpc$, in the two-loop evaluation.

Note that in Eq.~\eqref{eq:P_l_2L_ren} we defined EFT parameters independently
for each multipole, in contrast to typical conventions in the EFT literature,
e.g.~\cite{DAmico:2019fhj, Chudaykin:2020aoj}, where the counterterms are
defined in redshift space with certain $\mu$-dependencies. Given the numerical
complexity of integrating the two-loop single-hard counterterm over $\mu$, we
opt to parametrize the counterterms by their projection onto Legendre
polynomials, avoiding the need to perform the integral repeatedly when fitting
the EFT parameters. This also allows us to treat each multipole separately. We
stress that the approaches are equivalent when considering multipoles up to
at most $\ell=4$, and simply related by a change of basis in $\mu$-space.

\subsection{IR-resummation}\label{sec:IRresum}

To properly model the BAO wiggles in the power spectrum we need to include the
effect of large displacements. These are not adequately modeled in SPT; however,
how to resum their impact on the BAOs is
well-understood~\cite{Eisenstein:2006nj, Crocce:2007dt, Senatore:2014via, Baldauf:2015xfa,
Vlah:2015zda, Blas:2016sfa, Ivanov:2018gjr}.
After separating out the BAO wiggles (w) from the broadband (nw) linear
spectrum in real space, $P^{\text{lin}} = \Pnw + \Pw$
(for the numerical analysis in Sec.~\ref{sec:numerics} we perform this
splitting using the algorithm of \cite{Hamann:2010pw,Chudaykin:2020aoj}),
one defines the IR damping factor~\cite{Blas:2016sfa}
\begin{equation}
    \Sigma^2(z) = \frac{4\pi}{3}
    \int_0^{k_s} \dd q\,
    \Pnw(q)
    \left[
        1 - j_0
        \left(
            \frac{q}{k_{\text{osc}}}
        \right)
        + 2 j_2
        \left(
            \frac{q}{k_{\text{osc}}}
        \right)
    \right]
    \,,
    \label{eq:IR_Sigma2}
\end{equation}
where $k_{\text{osc}} = h/(110~\Mpc)$ is the wavenumber of the BAO
oscillations, $k_s$ is the separation scale between short and long modes in the
resummation and $j_n$ is the spherical Bessel function. In redshift space we
need the additional damping factor~\cite{Ivanov:2018gjr}
\begin{equation}
    \delta\Sigma^2(z) = 4\pi
    \int_0^{k_s} \dd q\,
    \Pnw(q) \,
    j_2
    \left(
        \frac{q}{k_{\text{osc}}}
    \right)
    \,,
    \label{eq:IR_delta_Sigma2}
\end{equation}
yielding the following total, anisotropic damping factor in redshift space (related to Eq.~\eqref{eq:Zsoft}):
\begin{equation}
    \Sigma_{\text{tot}}^2(z,\mu) =
    (1 + f(z)\mu^2 (2 + f(z))) \Sigma^2(z) +
    f^2(z)\mu^2(\mu^2 - 1)\delta\Sigma^2(z)
    \,.
    \label{eq:IR_Sigma2_tot}
\end{equation}
The IR-resummed perturbative contributions to the power spectrum then reads
\begin{subequations}
    \begin{align}
        P_{s,\text{LO}}^{\text{IR}}(k,\mu) &=
            \left( Z_1(k,\mu) \right)^2
            \left( \Pnw(k) + \e^{-k^2 \Sigma_{\text{tot}}^2(\mu)} \Pw(k) \right)
            \,,
        \\
        P_{s,\text{NLO}}^{\text{IR}}(k,\mu) &=
            \left( Z_1(k,\mu) \right)^2
        \left(
            \Pnw(k) +
            \left( 1 + k^2\Sigma_{\text{tot}}^2(\mu) \right)
            \e^{-k^2 \Sigma_{\text{tot}}^2(\mu)}
            \Pw(k)
        \right)
        \nonumber\\
        & \phantom{=} + P_{s}^{\text{1-loop}}
        \left[
            \Pnw(q) + \e^{-q^2 \Sigma_{\text{tot}}^2(\mu)} \Pw(q)
            \right]\,,
        \\
        P_{s,\text{NNLO}}^{\text{IR}}(k,\mu) &=
            \left( Z_1(k,\mu) \right)^2
        \left(
            \Pnw(k) +
            \left(
            1 + k^2\Sigma_{\text{tot}}^2(\mu) +
            \frac{1}{2} \left(k^2\Sigma_{\text{tot}}^2(\mu)\right)^2
            \right)
            \e^{-k^2 \Sigma_{\text{tot}}^2(\mu)}
            P_w(k)
        \right)
        \nonumber \\
        & \phantom{=} + P_{s}^{\text{1-loop}}
        \left[
            \Pnw(q) +
            \left( 1 + q^2\Sigma_{\text{tot}}^2(\mu) \right)
            \e^{-q^2 \Sigma_{\text{tot}}^2(\mu)}
            \Pw(q)
        \right]
        \nonumber \\
        & \phantom{=} + \bar P_{s}^{\text{2-loop}}
        \left[
            \Pnw(q) + \e^{-q^2 \Sigma_{\text{tot}}^2(\mu)} \Pw(q)
        \right]
        \,,
    \end{align}%
    \label{eq:P_s_IR}%
\end{subequations}
where the notation $P_{s}^{L\text{-loop}}[X]$ indicates that $X$ is the power
spectrum entering the integrand of the loop corrections. Note that
the additional terms $\sim k^2 \Sigma^2$ and $\sim (k^2 \Sigma^2)^2$ correct
for the overcounting of IR contributions already included in the loop integrals.

In summary, the renormalized and IR-resummed two-loop power spectrum in redshift
space is
\be
  P_{l,\text{NNLO}}^{\text{IR,ren.}}(k) =
  \frac{2l + 1}{2}
    \int_{-1}^{1} \dd\mu \, \mathcal{P}_l(\mu)
    \, P_{s,\text{NNLO}}^{\text{IR}}(k,\mu)
    +     \gamma_1(\Lambda) k^2 P_{l}^{\text{lin}}(k) +
    \gamma_2(\Lambda) \bar{P}_{l,\text{sh}}^{\text{2-loop}}(k;\Lambda)\,,
    \label{eq:P_l_L_ren}
\ee
where the last two terms are the EFT corrections discussed above, but evaluated
using $P_0(q) = \Pnw(q) + \e^{-q^2 \Sigma^2(\mu)} \Pw(q)$ as input spectrum.
The NLO prediction is computed analogously, but including only the $\gamma_1$ term.

\section{Numerical analysis}
\label{sec:numerics}

In this section we compare the two-loop matter power spectrum in redshift space, including EFT corrections and IR-resummation, to numerical $N$-body
simulation results from the Quijote suite~\cite{Villaescusa-Navarro:2019bje}.
The simulations feature 500 realizations of $(512)^3$ CDM particles in a cubic box of size $(1~\mathrm{Gpc}/h)^3$ with pair-fixed initial conditions which have reduced cosmic variance.
The cosmological parameters are $\Omega_m = 0.3175$, $\Omega_b = 0.049$, $h = 0.6711$, $n_s = 0.9624$ and $\sigma_8 = 0.834$.
We use the EdS-approximation for the matter kernels (see e.g.\ \cite{Garny:2020ilv, Garny:2022fsh} for analyses
of the impact of the EdS-approximation on the two-loop power spectrum in real
space). The loop integrals are cutoff at $\Lambda = 1\ihMpc$, however, as we check
explicitly in App.~\ref{app:cutoff}, the EFT prescription ensures that our
results do not depend on this choice.
We present results  for the monopole $\ell=0$ and quadrupole $\ell=2$.
A presentation of the hexadecapole $\ell=4$ is left to
Appendix~\ref{app:hexadecapole} due to it not being estimated as well by the
simulations.
Furthermore we consider redshifts $z=0$ and $z=0.5$ in Sec.~\ref{sec:z0}
and Sec.~\ref{sec:z0p5}, respectively.

We determine the EFT coefficients $\gamma_1$ and $\gamma_2$ defined in Eq.~\eqref{eq:P_l_L_ren}
by fitting the perturbative result to the corresponding simulation output, taking all $k$-bins up to some maximal wavenumber $k_\text{max}$ into
account. In practice, we minimize the function
\be
  \chi^2(k_\text{max}) \equiv \sum_{k_i\leq k_\text{max}} \frac{1}{\sigma_{k_i}^2}\left(P_{l,\text{NNLO}}^{\text{IR,ren.}}(k_i)-P_l^\text{Quijote}(k_i)\right)^2\,,
  \label{eq:chi2}
\ee
where $\sigma_{k_i}^2$ is determined from the variance of the 500 realizations provided in the Quijote simulation suite.
Throughout the analysis, the first $k$-bin of the Quijote redshift-space data vector is removed, which due to discreteness effects is inaccurate.
We reiterate that we perform the EFT fit independently for each multipole.

In order to compare our two-loop results to those at one-loop, we consider various different possibilities for the latter, varying in the number of free EFT parameters
allowed in the fit. The EFT schemes can by summarized as follows:
\begin{align*}
    &\text{NLO}                                      && \{\gamma_1\}                  && \text{one-loop, 1-parameter,}                                                                        \\
    &\text{NLO},~\gamma_1 = \gamma_1^{\text{[NNLO]}} && \{\emptyset\}                 && \text{one-loop, 0-parameter with } {\gamma}_1 \text{ fixed from NNLO fit,}                       \\
    &\text{NLO}+k^4                                  && \{\gamma_1, \gamma_{k^4} \}   && \text{one-loop, 2-parameter, including proxy counterterm } \gamma_{k^4} k^4 P_l^{\mathrm{lin}}(k)\,, \\
    &\text{NNLO}                                     && \{\gamma_1, {\gamma}_2 \} && \text{two-loop, 2-parameter\,.}
\end{align*}
The first scheme is the typical one-loop EFT prescription, which we find to be prone to overfitting already for moderate values of $k_\text{max}$ due to the limited
range of validity of the one-loop approximation. Following~\cite{Garny:2022fsh}, we therefore also present NLO results for which the leading EFT coefficient $\gamma_1$
is taken from the best-fit of the NNLO result. Alternatively, one could limit the range of $k_\text{max}$ at NLO compared to NNLO, but we opt for the former possibility
in order to exhibit the dependence on $k_\text{max}$ at NLO as compared to NNLO. Finally, in Sec.~\ref{sec:k4}, we also consider the option of adding a second EFT term to the
NLO result, as this scheme has been used in various analyses~\cite{Lewandowski:2015ziq, Ivanov:2019pdj, Nishimichi:2020tvu, DAmico:2021ymi}.

\subsection{Redshift $z = 0$}\label{sec:z0}

\begin{figure}[t]
    \centering
    \includegraphics[width=\textwidth]{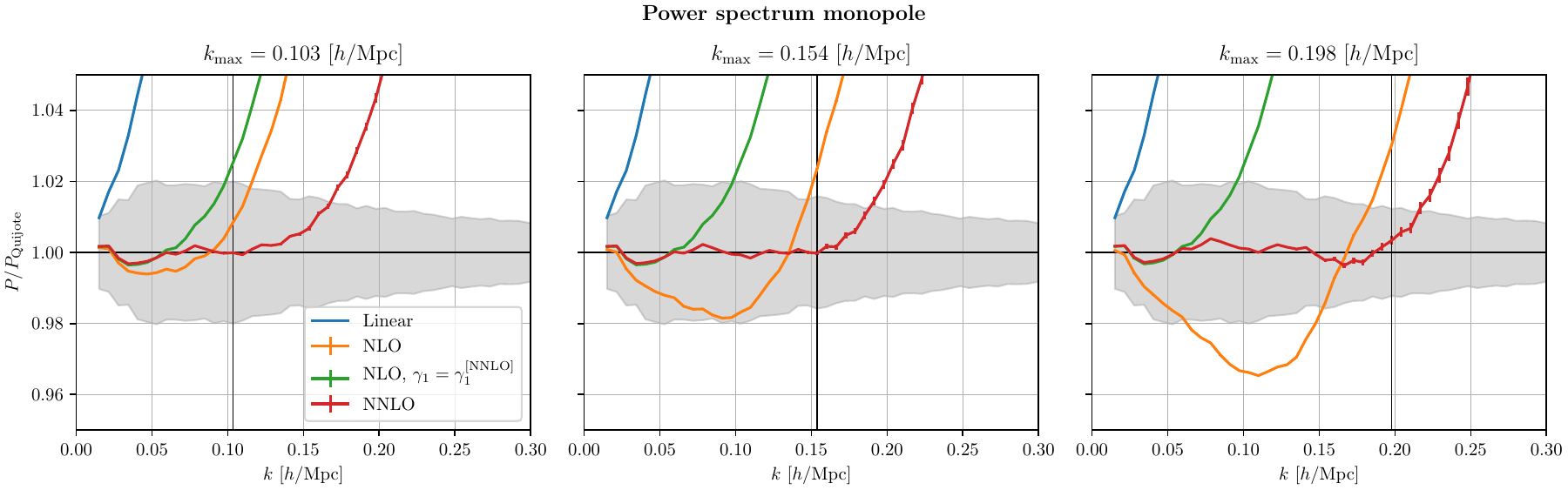} \\[0.3cm]
    \includegraphics[width=\textwidth]{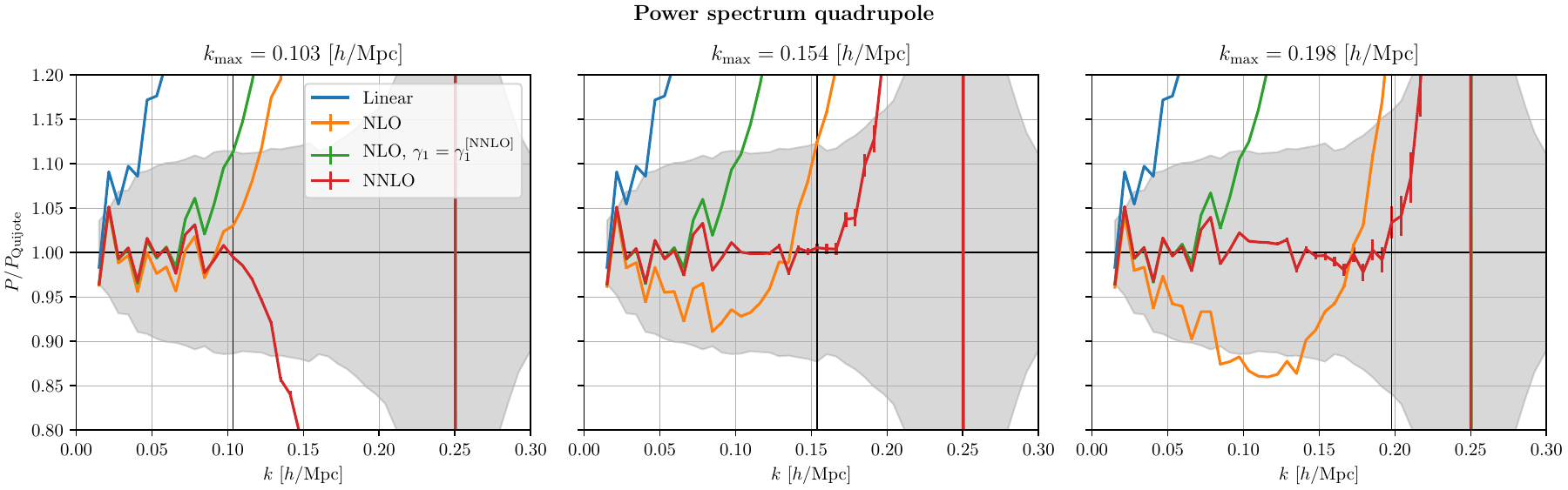}
    \caption{Predictions for the monopole (upper panel) and quadrupole (lower
        panel) contributions to the power spectrum in redshift space at $z =
        0$, normalized to the Quijote result. We display the linear (blue), NLO
        0-parameter (green), NLO 1-parameter (orange) and NNLO (red)
        predictions using three pivot scales $k_\text{max}\simeq 0.1$, $0.15$
        and $0.2\ihMpc$ from left to right. The Quijote uncertainty
        due to sample variance is indicated by the gray shading, and errorbars on
        the theory predictions mark the uncertainty from the numerical loop
        integration.}
    \label{fig:pk_l02_z0}
\end{figure}

\begin{figure}[ht]
    \centering
        \includegraphics[width=0.8\textwidth]{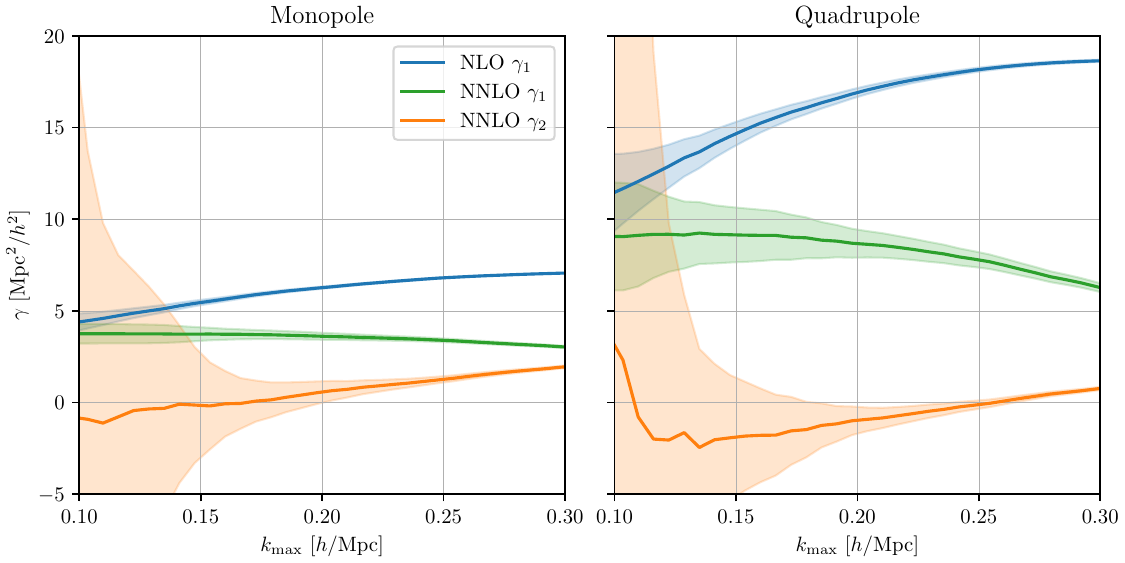}
        \caption{Measured EFT parameters as a function of pivot scale $\kmax$,
        within the 1-parameter NLO (blue) and 2-parameter NNLO (green and orange) EFT models,
        for the monopole (left) and quadrupole (right), all at $z=0$.
        Shaded regions indicate $1\sigma$ uncertainty from the fit.}
    \label{fig:gamma_z0}
\end{figure}

Our result for the NNLO power spectrum in redshift space  at $z=0$  is shown in Fig.~\ref{fig:pk_l02_z0} for the monopole (upper row)
and quadrupole (lower row), and three values of $k_\text{max}$, respectively. The power spectra are normalized to those obtained from the Quijote simulation,
with gray shaded areas bracketing its uncertainty. As expected, the simulation results are subject to larger statistical fluctuations for the quadrupole,
with its relative size becoming particularly large close to the scale where it changes sign at around $0.25\ihMpc$.

For our perturbative analysis we limit ourselves to the regime below  both the FoG and non-linear scales.
The former can be identified with the scale for which $k^2f^2\sigma_v^2=1$, where $\sigma_v^2\equiv 1/3\int_{\vq} P_{\theta\theta}(q)/q^2$
is related to the variance of the velocity power spectrum~\cite{Scoccimarro:2004tg}. Using the linear spectrum to compute the variance
yields $k_{\text{FoG}}(z = 0) = 0.198\ihMpc$ for the considered cosmology. As well-known this scale is somewhat below the usual non-linear
scale, $k_{\text{NL}}(z = 0)   = 0.31\ihMpc$, defined by the wavenumber for which the variance of the (linear) density field smoothed on that scale
is equal to unity.

We find that the NNLO power spectrum is in very good agreement with Quijote results for all $k\lesssim k_\text{max}$ for both the monopole and
quadrupole when choosing $k_\text{max}\simeq 0.1$ or $0.15\ihMpc$ (red lines in left and middle columns in Fig.~\ref{fig:pk_l02_z0}, respectively). The agreement extends
to wavenumbers somewhat above the pivot scale particularly for the monopole. For the largest pivot scale $k_\text{max}\simeq 0.2\ihMpc$ considered in our analysis (red line in right column in Fig.~\ref{fig:pk_l02_z0}) we observe a rapid degradation of the agreement between NNLO and simulation results for $k>0.2\ihMpc$, in accordance with the approach of the FoG scale.
Nevertheless, the EFT model is still able to describe the Quijote data at smaller wavenumbers for both the monopole and quadrupole. Overall, taking the middle pivot scale as fiducial choice,
we find agreement within sample uncertainty from the Quijote data up to around $k\leq 0.18\ihMpc$ at NNLO.

In contrast, the range of validity of the corresponding NLO predictions is more limited. For the standard 1-parameter NLO model, it is obvious that the fit to simulation data
leads to a significant scale-dependent deviation for $k_\text{max}\simeq 0.15$ and $0.2\ihMpc$, that we interpret as a sign of overfitting (orange lines in middle and right columns in Fig.~\ref{fig:pk_l02_z0}, respectively). For comparison we therefore also display the 0-parameter NLO model described above, for which the EFT parameter $\gamma_1$ is taken from the NNLO fit result (green lines). Due to the larger range of validity of the NNLO prediction, the potential impact of overfitting is mitigated in this way. We therefore consider the 0-parameter NLO result more appropriate for large $k_\text{max}$. For the smallest pivot scale $k_\text{max}\simeq 0.1\ihMpc$, the 0- and 1-parameter NLO results are more close to each other, signaling that overfitting is
less of an issue for this choice (orange and green lines in left column of Fig.~\ref{fig:pk_l02_z0}). When excluding the setups that are obviously affected by overfitting, the range of agreement between NLO and Quijote within uncertainties lies around $0.08-0.12\ihMpc$ depending on the choice of pivot scale, multipole and 0- versus 1-parameter model. For an unbiased comparison to NNLO we adopt the same pivot scale $k_\text{max}\simeq 0.15\ihMpc$ and the 0-parameter model as fiducial choice, for which we find agreement between NLO and Quijote data for $k\leq 0.1\ihMpc$.

In addition to the power spectra, it is also instructive to investigate the dependence of the best-fit EFT parameters on $k_\text{max}$, shown in Fig.~\ref{fig:gamma_z0}.
Within the validity range, one expects the EFT parameters to be approximately independent of $k_\text{max}$, which is the case at NNLO for pivot wavenumbers below the FoG scale (green and orange lines). Interestingly, the second EFT parameter $\gamma_2$ is almost compatible with being zero over a wide range. Note that it can only be determined reliably when going to moderately large $k_\text{max}$ values, since the corresponding single-hard EFT correction is highly suppressed for low $k$ values, explaining the large
uncertainty region for $\gamma_2$ in Fig.~\ref{fig:gamma_z0}. The $\gamma_1$ parameter is significantly larger as compared to the
real space power spectrum~\cite{Baldauf:2015aha} (see also App.~\ref{app:real}). This may be expected due to EFT corrections setting in at the FoG instead of the non-linear scale. Finally, we observe a significant
scale-dependence of $\gamma_1$ obtained within the 1-parameter NLO model, particularly for the quadrupole (blue lines in Fig.~\ref{fig:gamma_z0}). This can be taken as another indication that the validity range of the 1-parameter NLO result is rather limited.

Altogether, we find that going from NLO to NNLO extends the range of agreement
with Quijote simulation data including sampling uncertainty for the monopole
and quadrupole from $k\simeq 0.1\ihMpc$ to $k\simeq 0.18\ihMpc$ at $z=0$.

\subsection{Redshift $z = 0.5$}\label{sec:z0p5}

\begin{figure}[t]
    \centering
    \includegraphics[width=\textwidth]{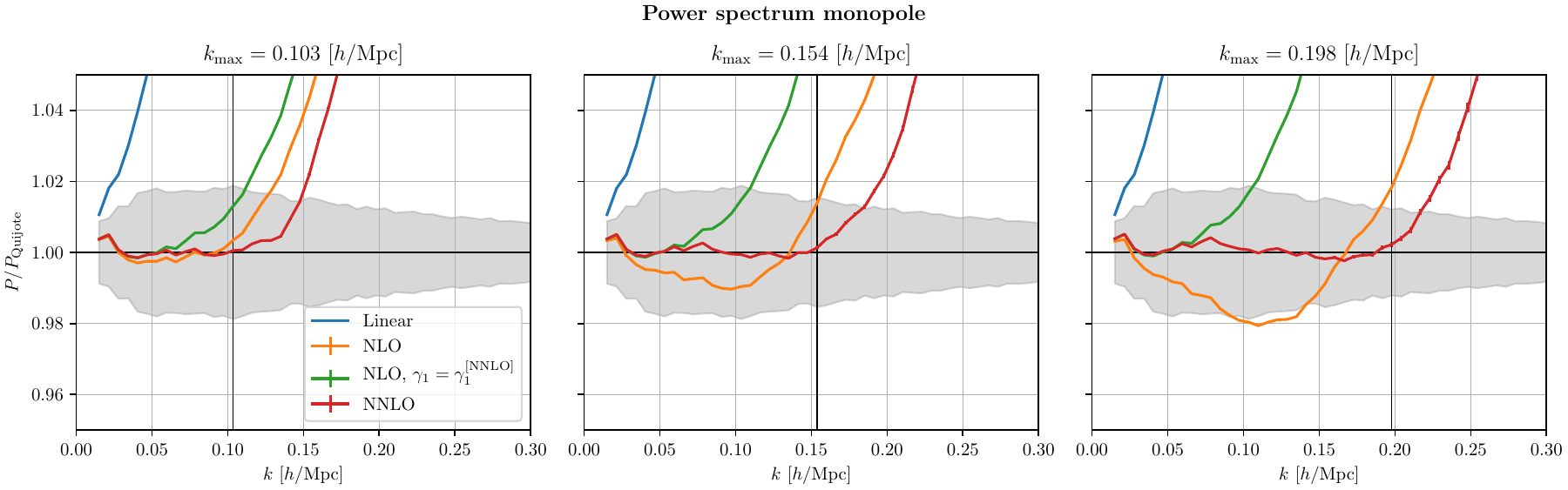} \\[0.3cm]
    \includegraphics[width=\textwidth]{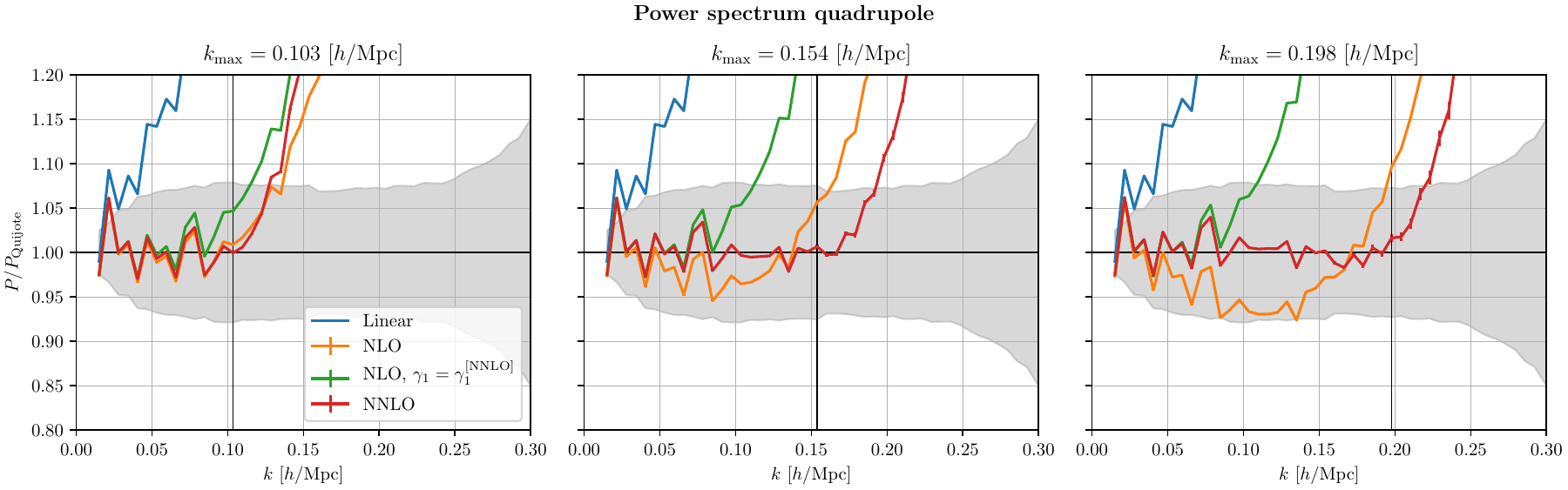}
    \caption{Same as Fig.~\ref{fig:pk_l02_z0} for redshift $z = 0.5$.}
    \label{fig:pk_l02_z0p5}
\end{figure}

For completeness, we show the NLO and NNLO results at redshift $z=0.5$ in Fig.~\ref{fig:pk_l02_z0p5}, for the same analysis choices as in the previous section.
As is well-known, the FoG scale is almost unchanged compared to $z=0$, $k_{\text{FoG}}(z = 0.5) = 0.220\ihMpc$, due to the counteracting effect of a smaller growth factor $D(a)$ but a larger growth rate $f(a)$. Thus, even though the non-linear scale $k_{\text{NL}}(z = 0.5) = 0.47\ihMpc$ is around 50\% larger compared to $z=0$, a similar range of validity of the perturbative EFT model is expected in redshift space at both redshift values.

Our numerical results largely confirm this expectation, and feature similar properties as observed for $z=0$. In particular, the NNLO result shows very good agreement with Quijote data up to the respective pivot scales, while the 1-parameter NLO models suffers from overfitting for $k_\text{max}\simeq 0.15$ and $0.2\ihMpc$. Adopting therefore the 0-parameter NLO model for comparison, and the same fiducial pivot scale as at $z=0$ (i.e.\ the middle one), we find agreement within Quijote uncertainty for the monopole and quadrupole up to $0.11\ihMpc$ at NLO, and up to $0.18\ihMpc$ at NNLO.

\subsection{Add $k^4$ counterterm to NLO prediction}\label{sec:k4}

\begin{figure}[t]
    \centering
        \includegraphics[width=\textwidth]{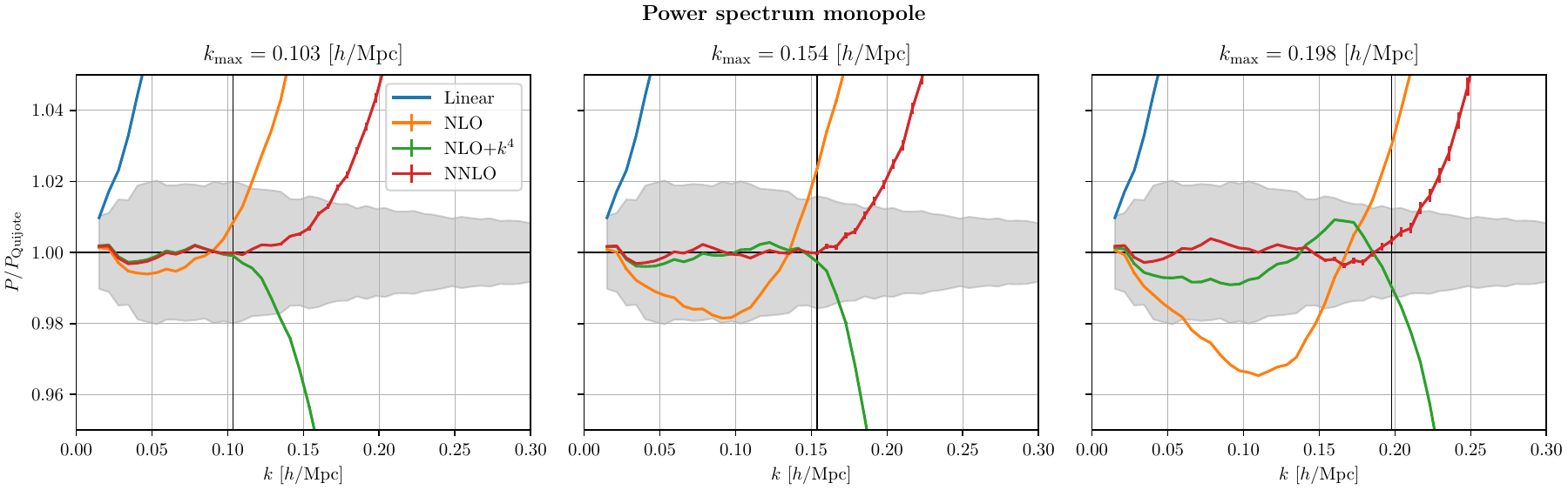}
        \includegraphics[width=\textwidth]{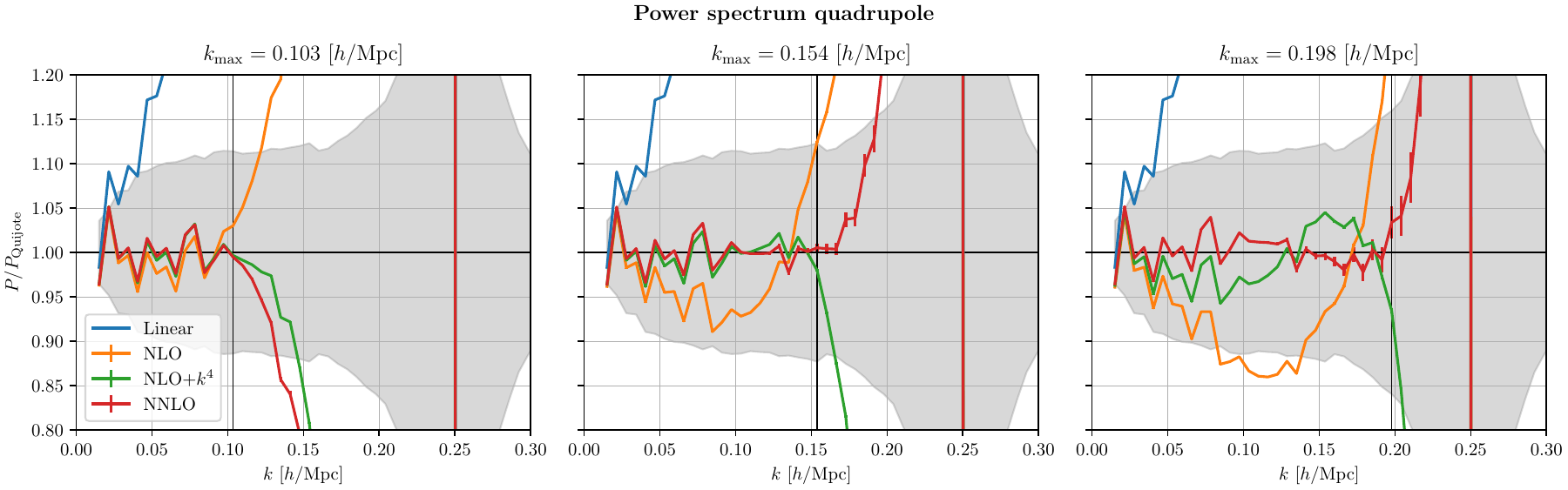}
        \caption{Impact of introducing a proxy $k^4P_l^{\mathrm{lin}}(k)$
            counterterm at NLO. Theoretical predictions at $z = 0$ for the
            monopole and quadrupole normalized to Quijote $N$-body data are shown. The
            green line corresponds to the NLO prediction including a $k^4
            P_l^{\mathrm{lin}}(k)$ counterterm (2 free parameters in total), and
            the 1-parameter NLO as well as the NNLO results are displayed for
            comparison. Gray shadings indicate Quijote uncertainty, and
            errorbars mark uncertainty from numerical loop integration.}
    \label{fig:pk_l0l2_k4}
\end{figure}

To account for the FoG effect within the EFT setup, it has been argued that the one-loop result for the power spectrum in redshift space can be
improved by allowing for an extra additive term scaling as the product of $k^4$ times the linear spectrum, which can be viewed as a higher-derivative
correction to the common $k^2$ term~\cite{Lewandowski:2015ziq, Ivanov:2019pdj, Nishimichi:2020tvu}.
In order to assess the impact of adding this correction at NLO as compared to our NNLO results,
we consider a 2-parameter NLO model in this section. In particular, relative to the 1-parameter NLO power spectrum we allow for
an additional correction term $k^4P_l^\text{lin}(k)$, with its prefactor being a second free EFT parameter for each multipole.
We note that this is slightly more general than the template from~\cite{Ivanov:2019pdj, Nishimichi:2020tvu}, where the magnitude
of this correction for the monopole is correlated to that for the quadrupole. This generalization is justified within
the EFT context as one expects several corrections at order $k^4$, differing in their $\mu$-dependence. We thus treat the EFT corrections to the
 monopole and to the quadrupole independently from each other in our analysis.

We show the result for the 2-parameter NLO model in Fig.~\ref{fig:pk_l0l2_k4} (green lines), and also the previous 1-parameter NLO (orange)
and 2-parameter NNLO (red) for comparison. Adding the $k^4$ term notably improves the agreement of the NLO prediction with Quijote data, especially at
pivot scales $k_\text{max}\simeq 0.1$ and $0.15\ihMpc$. However, for $k_\text{max}\simeq 0.2\ihMpc$ the NNLO result is markedly closer to Quijote than the 2-parameter NLO fit,
indicating that the latter is entering the overfitting regime. Nevertheless, it is interesting that the improvement from including two-loop corrections can be somewhat mimicked
by the 2-parameter NLO model for $k\lesssim 0.15\ihMpc$. When using scales beyond that and aiming at (sub-)percent accuracy, the inclusion of the NNLO result is still warranted.

Finally, one may wonder about the impact of adding an extra $k^4$-term at NNLO, which would then feature three EFT parameters. We checked that this does not yield any improvement
as compared to our fiducial 2-parameter NNLO result, and is therefore an unnecessary complication (see App.~\ref{app:k4} for details).

\section{Conclusion}
\label{sec:conclusion}

In this work we presented the matter power spectrum in redshift space including two-loop corrections.
We followed a strictly perturbative approach complemented by EFT corrections, that can in principle
be expected to be applicable up to scales $k\lesssim 0.2\ihMpc$ beyond which non-perturbative FoG suppression sets in.
In our work we extended the two-loop scheme proposed in~\cite{Baldauf:2015aha} from real to redshift space.
In particular, we add an extra EFT correction term for each multipole moment of the power spectrum, which accounts for
the UV-sensitivity of the two-loop contribution when one of the loop wavenumbers becomes large (single-hard limit).
The double-hard limit can be renormalized by the same $k^2P_l^\text{lin}(k)$ correction included already at one-loop, as expected.
In total, this leads to two free parameters for each of the multipoles we considered, specifically the monopole and quadrupole.

We performed various numerical and analytical checks to confirm the validity of the setup, and incorporated also IR-resummation.
The NNLO result was calibrated with and compared to Quijote $N$-body simulation data for $z=0$ and $z=0.5$. At NNLO we find that
the EFT parameters are stable against variation of the maximal wavenumber included in the fit for $k_\text{max}\lesssim 0.2\ihMpc$,
showing no sign of overfitting. In contrast, when using the NLO result to determine the $k^2P_l^\text{lin}(k)$ correction,
we find a strong scale-dependence of the best-fit coefficient within the range $k_\text{max}\gtrsim 0.1\ihMpc$ allowed by $N$-body data.
In addition, the $k$-dependence of the NLO result determined in this way shows indications of overfitting.
We therefore opted to use the $k^2P_l^\text{lin}(k)$ correction determined at NNLO for our fiducial NLO setup as well. We then find that the NLO result is in agreement
with Quijote data for the monopole and quadrupole at $z=0$ until $k\simeq 0.1\ihMpc$. Adding the NNLO correction significantly extends the
range of agreement to $k\simeq 0.18\ihMpc$. We find a similar improvement from $k\simeq 0.11\ihMpc$ at NLO to $k\simeq 0.18\ihMpc$ at NNLO for redshift $z=0.5$.
This is consistent with the FoG scale being around $0.2\ihMpc$ for both redshifts within the $\Lambda$CDM cosmology adopted in this work.

Finally, we investigated also the option proposed in~\cite{Ivanov:2019pdj} to add a higher-derivative correction. We considered terms of the
form $k^4P_l^\text{lin}(k)$, with an extra free parameter for each multipole, being slightly more general than the ansatz from~\cite{Ivanov:2019pdj}.
At NNLO, adding this higher-derivative term does not lead to any changes compared to our fiducial setup, and we therefore omit it. However, at NLO this
additional freedom can bring the result closer to $N$-body data for $k\lesssim 0.15\ihMpc$. Nevertheless, especially when pushing to scales $k_\text{max}\simeq 0.2\ihMpc$ the
fiducial NNLO result is still in very good agreement with Quijote data while the NLO result deviates significantly even when allowing for two free parameters ($k^2P_l^\text{lin}(k)$ and
 $k^4P_l^\text{lin}(k)$ terms) for each multipole for the latter.

We provide details on our implementation of redshift-space kernels as well as complementary results, including analytical expressions for the two-loop EFT correction to the
monopole and quadrupole power spectrum, in the appendices, intended for convenient usage in future works.

\subsection*{Acknowledgements}
We thank Roman Scoccimarro for insightful discussions and Francisco Villaescusa-Navarro and Oliver Philcox for helpful comments
on using the redshift-space power spectra from the Quijote simulation suite.
We acknowledge support by the ANR Project COLSS (ANR-21-CE31-0029, France)
and by the DFG Collaborative Research Institution Neutrinos and Dark Matter in
Astro- and Particle Physics (SFB 1258).

\appendix

\section{Redshift-space kernels}
\label{app:kernels}

In this appendix we write down a general formula for the redshift-space kernels
and briefly discuss the algorithm we employ to evaluate it for a given set of
wavevectors. At $n$-th order, the redshift-space kernel (defined implicitly in
Eq.~\eqref{eq:Z_n}) can be written as
\begin{equation}
    Z_n(\vq_1,\dotsc,\vq_n) =
    A_n(\vq_1,\dotsc,\vq_n) +
    \sum_{m=1}^{n-1}
    A_m(\vq_1,\dotsc,\vq_m)
    \sum_{p=1}^{n-m}
    \frac{(f\mu k)^p}{p!}
    B_{n-m}^{(p)}(\vq_{m+1},\dotsc,\vq_n)
    \,,
    \label{eq:Z_n_general}
\end{equation}
where $\vk = \sum_{i=1}^{n} \vq_i$ and $\mu = \vk \cdot \hat{z}/k$. We defined
$A_m$ as a shorthand notation for the $m$-th order perturbative expansion of
the square bracket of the first line in Eq.~\eqref{eq:Z_n}:
\begin{equation}
    A_m(\vq_1,\dotsc,\vq_m) =
    F_m(\vq_1,\dotsc,\vq_m) +
    f \mu_{1\dotsc m}^2
    G_m(\vq_1,\dotsc,\vq_m)
    \,,
\end{equation}
where $\mu_{1\dotsc m} = \vq_{1\dotsc m} \cdot \zhat / q_{1\dotsc m}$ and
$\vq_{1\dotsc m} = \sum_{i=1}^{m} \vq_i$. The $p$-sum counts contributions with
$p$ factors of velocity kernels (arising from the expansion of the exponential
in Eq.~\eqref{eq:delta_s}). Hence, $B_m^{(p)}$ is defined as the sum of all
such terms at $m$-th order in perturbation theory with $p$ factors of velocity
kernels, which can more precisely be written recursively as
\begin{equation}
    B_m^{(p)}(\vq_1,\dotsc,\vq_m) =
    \begin{dcases}
        \sum_{l=p-1}^{m-1}
        B_l^{(p-1)}(\vq_1, \dotsc, \vq_l) \, B_{m-l}^{(1)}(\vq_{l+1}, \dotsc, \vq_m)
            \,, \quad & p > 1\,, \\
        \frac{\mu_{1\dotsc m}}{q_{1\dotsc m}} \, G_m(\vq_1,\dotsc,\vq_m)
            \,, \quad & p = 1\,.
    \end{dcases}
    \label{eq:B_n}
\end{equation}

Finally, to obtain a kernel $Z_n$ that is symmetric in its arguments, we need
$A_n$ and $B_n^{(p)}$ symmetric and the sum over $m$ in
Eq.~\eqref{eq:Z_n_general} must be symmetrized with respect to all permutations
exchanging momenta from the $\{\vq_1,\dotsc,\vq_m\}$ set with the
$\{\vq_{m+1},\dotsc,\vq_n\}$ set. $A_n$ is trivially symmetric provided that
the standard $F_n$ and $G_n$ kernels are; we can make $B_n^{(p)}$ symmetric by
symmetrizing the sum over $l$ in Eq.~\eqref{eq:B_n} the same way.

We use Eq.~\eqref{eq:Z_n_general} to implement an algorithm that evaluates the
redshift-space kernels for a given set of momenta. Using the algorithm of
Ref.~\cite{Blas:2013aba} we can efficiently obtain $F_n$ and $G_n$ and thus
$A_n$. We implement $B_n^{(p)}$ recursively utilizing memorization: we store
results on a grid defined by $n$, $p$ as well as an integer representing the
momentum argument configuration, with a hash-function for easy lookup.
Storing intermediate results greatly speeds up the algorithm due to the fact
that the same combinations of $n$, $p$ and argument configurations of
$B_n^{(p)}$ appears several times in the recursive sum of Eq.~\eqref{eq:B_n},
the double sum of Eq.~\eqref{eq:Z_n_general} and additionally because of
overlap between the different kernels required to compute loop corrections to
the power spectrum (cf.\ Eq.~\eqref{eq:integrand}). Both authors independently
implemented the algorithm in separate codes and checked that the calculated
one- and two-loop corrections agree within Monte Carlo numerical uncertainty
between the codes.

\section{Additional checks and supplementary results}
\label{app:validation}

In this appendix we perform various tests to validate our EFT framework at
NNLO, and provide supplementary results. We check that the
low-$k$ limit of the numerical evaluation at two-loop has the appropriate
scaling as required by mass and momentum conservation in Sec.~\ref{app:subtraction}.
Next, in Sec.~\ref{app:cutoff} we confirm numerically that the EFT prescription has
removed the cutoff-dependence of the bare theory by comparing EFT predictions
with different cutoffs. We apply our method in real space in Sec.~\ref{app:real}. In
Sec.~\ref{app:chi2} we show the $\chi^2$ corresponding to the fits shown in
Sec.~\ref{sec:numerics}. In Sec.~\ref{app:k4} we add a proxy $k^4$
counterterm to the \emph{NNLO} results, and finally in
Sec.~\ref{app:hexadecapole} we show results for the hexadecapole in redshift space.

\subsection{Subtraction of double-hard contribution}
\label{app:subtraction}

\begin{figure}
    \begin{center}
        \includegraphics[width=0.8\textwidth]{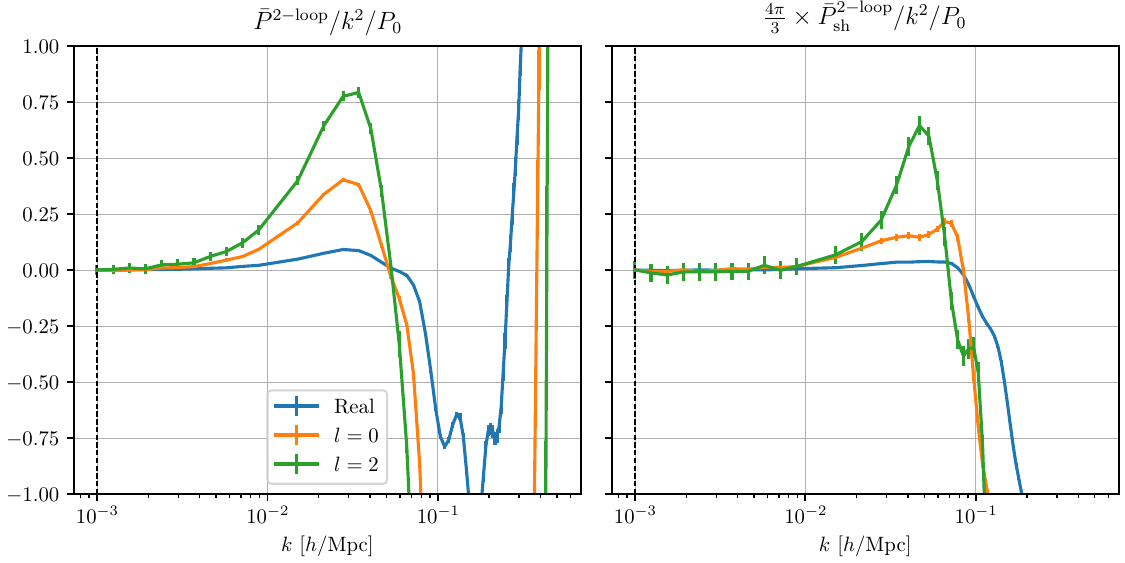}
    \end{center}
    \caption{Subtracted two-loop (\emph{left}) and single-hard two-loop
        (\emph{right}) corrections. The graphs are normalized by $k^2 P_0(k)$
        in order to highlight the double-hard scaling at low-$k$.
        For the real result, the spectrum $P_0$ used in the normalization stands for the IR-resummed input spectrum $\Pnw + \e^{-k^2
        \Sigma^2} \Pw$ in real space, and for the monopole and quadrupole it stands for the Kaiser spectrum computed with IR-resummed input spectrum $\Pnw + \e^{-k^2
        \Sigma^2_\text{tot}} \Pw$ in redshift space, projected onto the zeroth and second Legendre polynomials,
        respectively. The dashed line indicates the grid point at which the
        double-hard limit is fitted and removed. Errorbars correspond to numerical
        uncertainty from Monte Carlo integration.}
    \label{fig:two_loop_subtraction}
\end{figure}

As discussed in Sec.~\ref{sec:pt}, we choose a renormalization scheme where we
remove the double-hard contribution of the two-loop correction, so that the
$\gamma_1$ parameter is a priori expected to retain the same value at NLO and NNLO.
The double-hard limit is given analytically in
App.~\ref{sec:doublehardEFT}, however in practice we opt to measure this
limit by considering the low-$k$ limit of the two-loop correction. At $k\sim
10^{-3}\ihMpc$ the two-loop integral has only support for $q_1,q_2 \gg k$ (the
linear power spectrum is suppressed on large scales), corresponding to the
double-hard limit. Thus, we may fit and remove the double-hard $k^2 P_0(k)$
contribution. This is shown in the left panel of
Fig.~\ref{fig:two_loop_subtraction}. We display the \emph{subtracted} two-loop
correction, i.e.\ after the double-hard contribution has been removed. The
graphs are normalized to $k^2 P_0(k)$, which means that the subtraction
amounts to shifting them by a constant so that the low-$k$ limit goes to zero.
We see indeed that for $k\lesssim 0.004 \ihMpc$ the real result, the monopole
and the quadrupole all approach a plateau around zero, confirming that the
numerical result exhibits the correct double-hard limit, which can be
renormalized by the $\gamma_1$ counterterm.

Similarly, the right panel of Fig.~\ref{fig:two_loop_subtraction} displays the
subtracted single-hard limits of the two-loop corrections, normalized to
$k^2 P_0(k)$. Again, we observe a plateau at low-$k$, affirming the appropriate
scaling of the ``hard limit of the single-hard limit'' in the numerical result.
Compared to the full two-loop (right panel), the single-hard reaches the $k^2
P_0(k)$ scaling for even larger $k$, $k \lesssim 0.01 \ihMpc$, because $q_1$
is hard (in the numerics $q_1 = 10\ihMpc$).
We check that the $k$-scaling of the numerical single-hard result matches the analytic
results provided in App.~\ref{sec:singlehardEFT}.

\subsection{Cutoff-dependence}
\label{app:cutoff}

\begin{figure}
    \begin{center}
        \includegraphics[width=\textwidth]{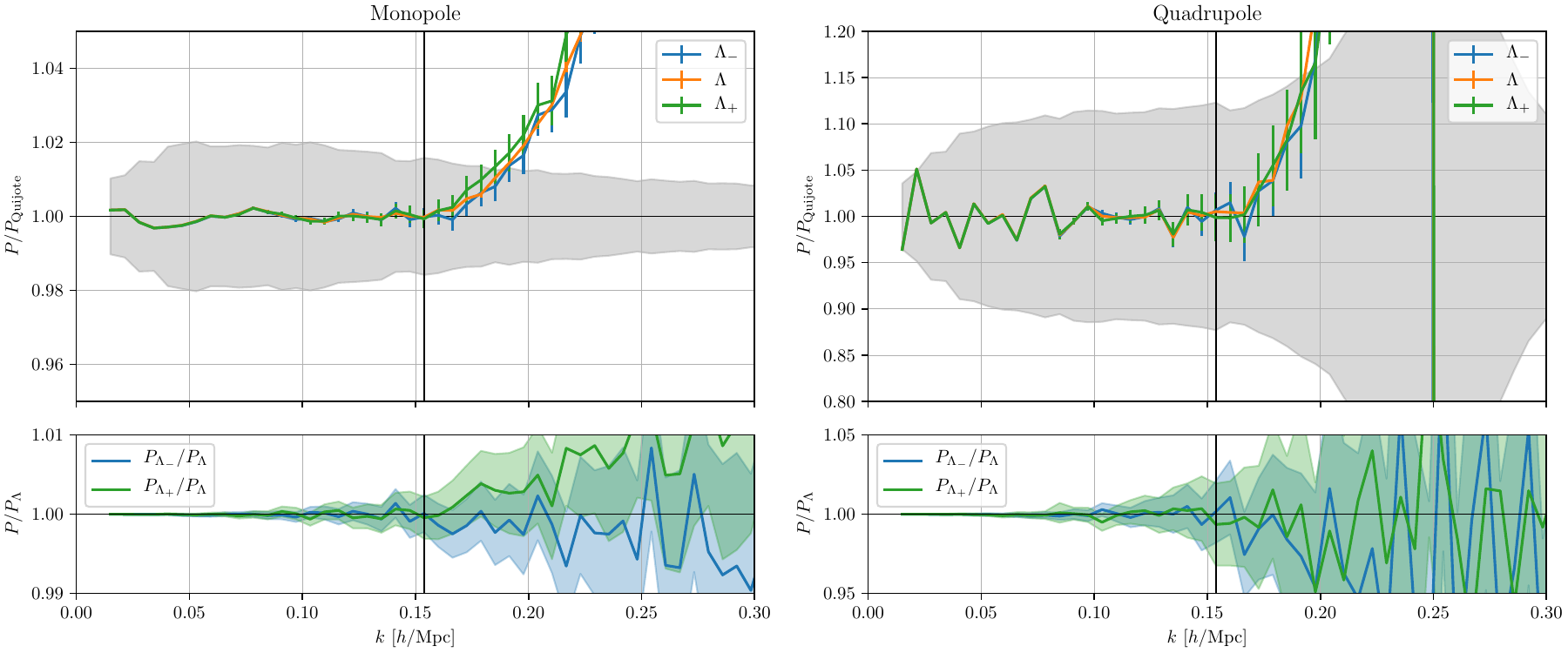}
    \end{center}
    \caption{Monopole and quadrupole contributions for different values of the
        cutoff: $\Lambda_{-} = 0.7\ihMpc$, $\Lambda = 1\ihMpc$ (our fiducial
        value) and $\Lambda_{+} = 1.3\ihMpc$. \emph{Upper panel:} predictions
        are normalized to Quijote, with the gray shaded region indicating
        simulation uncertainty. \emph{Lower panel:} fractional difference
        between the results using cutoffs $\Lambda_{-}$ and $\Lambda_{+}$ versus
        $\Lambda$.}
    \label{fig:cutoff}
\end{figure}

In this section we demonstrate that the renormalization in the EFT setup we
adopt indeed leads to a cutoff-independent prediction for the power spectrum in
redshift space, by computing results for two additional values of the cutoff to
the fiducial results shown in Sec.~\ref{sec:numerics}. The comparison is
displayed in Fig.~\ref{fig:cutoff}. In the upper panels we plot the monopole
(left) and quadrupole (right) power spectrum divided by the Quijote result for
three different cutoffs: $\Lambda_{-} = 0.7\ihMpc$, $\Lambda = 1\ihMpc$
(fiducial value used in main text) and $\Lambda_{+} = 1.3\ihMpc$. The lower
panels show the residuals between $\Lambda_{+}$/$\Lambda_{-}$ and $\Lambda$. We
see that the different cutoffs yield results that completely agree within
numerical uncertainty from the Monte Carlo integration, demonstrating that the
counterterms in Eq.~\eqref{eq:P_l_L_ren} appropriately correct for the
cutoff-dependence of the bare theory. The comparison is shown for a pivot scale
$\kmax = 0.15\ihMpc$, however we find the same conclusion for the other pivot
scales used in this work.

\subsection{Real space}
\label{app:real}

\begin{figure}
    \begin{center}
        \includegraphics[width=\textwidth]{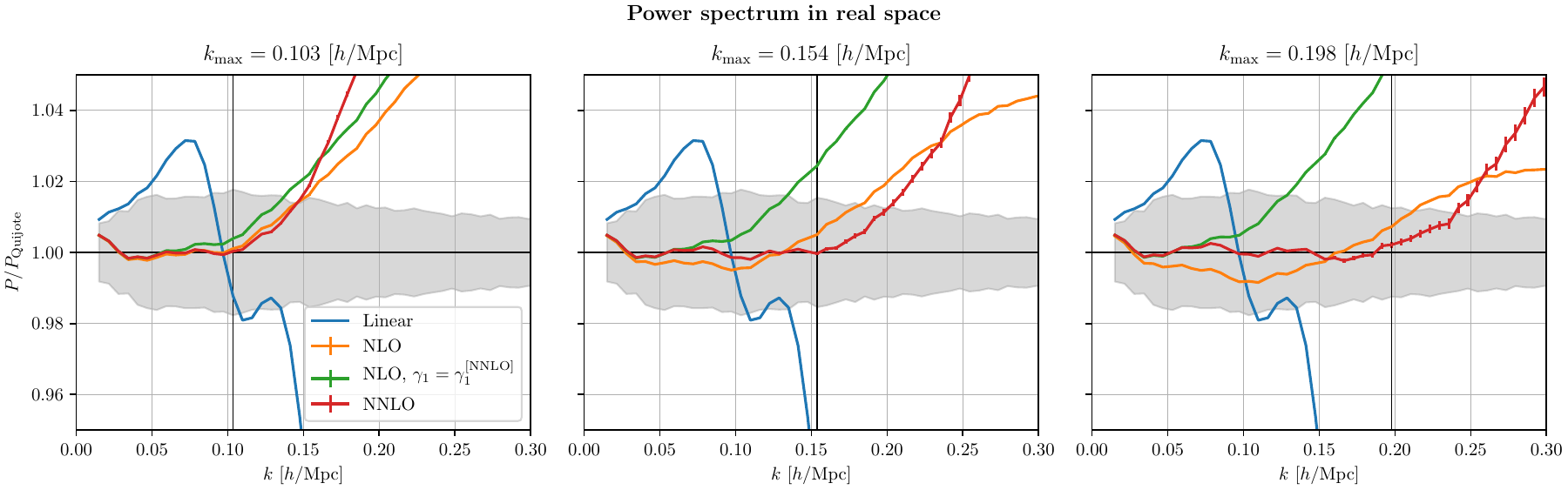}
    \end{center}
    \caption{Perturbative predictions for the power spectrum in real space at
        redshift $z=0$. The linear (blue), 0- (green) and 1-parameter (orange) NLO as well as NNLO (red) approximations are shown. The results are normalized to Quijote $N$-body data, and
        its uncertainty is displayed in gray.}
    \label{fig:pk_real}
\end{figure}

Next, we validate our method in real space. We use Eq.~\eqref{eq:P_2L_ren} and
the IR-resummation of Eq.~\eqref{eq:P_s_IR}, where in real space
$\Sigma_{\mathrm{tot}}^2 = \Sigma^2$. In Fig.~\ref{fig:pk_real} we show fits of
perturbative results in real space to Quijote $N$-body data for different pivot
scales. The calibration cases are the same as those presented in
Sec.~\ref{sec:numerics}: we show 0/1-parameters NLO as well as the
2-parameter NNLO results. As was the case in redshift space, if we use pivot
scales of $\kmax = 0.15\ihMpc$ or $0.2\ihMpc$, the standard 1-parameter NLO
model features spurious $k$-dependence that we interpret as overfitting. Hence,
we include also a 0-parameter NLO fit, using $\gamma_1$ as measured from the
NNLO fit, which should reduce the amount of overfitting and thus provide a
more faithful estimate for the wavenumber reach of the NLO prediction.
Using again $\kmax = 0.15\ihMpc$ as our fiducial pivot scale, we find that
adding the two-loop correction extends the wavenumber reach that matches
Quijote data within sampling uncertainty from $k = 0.13\ihMpc$ to $k=0.2\ihMpc$.
As expected, due to the smaller relative importance of higher-order corrections
in real space than in redshift space, perturbation theory in real space matches
$N$-body results down to somewhat smaller scales than in redshift space, although
the difference is not very large at $z=0$.

In total, we find that we can accurately match the Quijote $N$-body simulation
result to $k \simeq 0.2\ihMpc$ at $z=0$ with perturbation theory at NNLO,
using two EFT parameters, consistent with previous studies at
two-loop~\cite{Baldauf:2015aha, Garny:2022fsh}.

\subsection{Chi-squared}
\label{app:chi2}

\begin{figure}
    \begin{center}
        \includegraphics[width=0.8\textwidth]{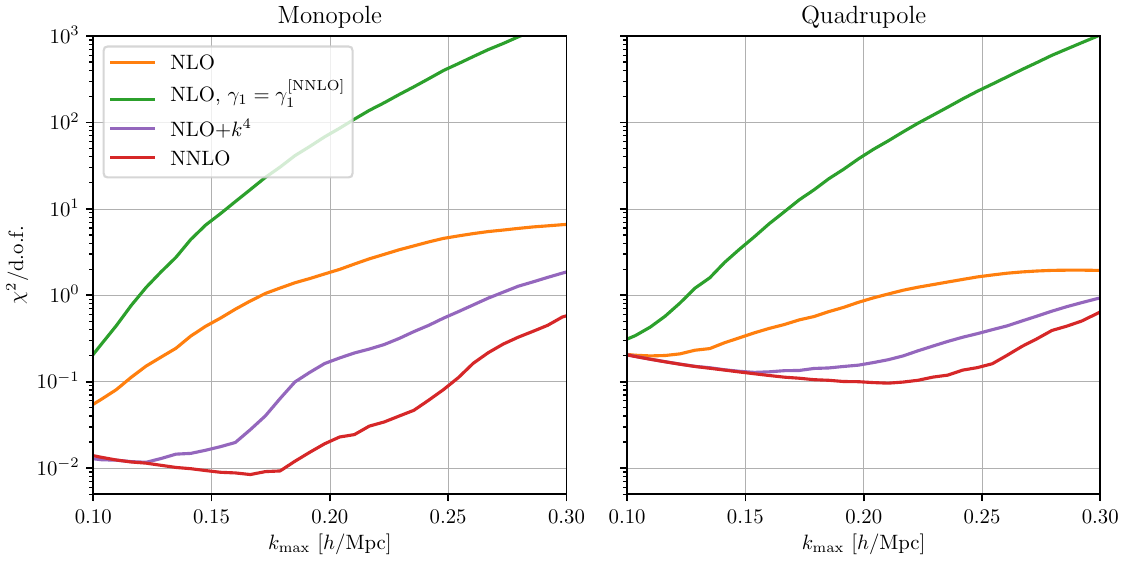}
    \end{center}
    \caption{Reduced $\chi^2$ for the various perturbative orders and
    calibration schemes (see Sec.~\ref{sec:numerics}), plotted as a function of the pivot scale
    $\kmax$.}
    \label{fig:chi2}
\end{figure}

The $\chi^2$ per d.o.f.\ (Eq.~\ref{eq:chi2}) at $z = 0$ is shown in
Fig.~\ref{fig:chi2} for the monopole/quadrupole and for the different
calibration cases discussed in Sec.~\ref{sec:numerics}. The small
$\chi^2/\text{d.o.f.} \ll 1$ suggests that our estimator for the $N$-body error
is too simple, e.g.\ not accounting for covariance of different bins, however
we are here only interested in the relative performance of each case, not the
overall $\chi^2$ value. As concluded in Sec.~\ref{sec:numerics}, the
0-parameter NLO fit (green) matches the Quijote data up to $k = 0.1\ihMpc$, and indeed
beyond this wavenumber we observe a steep increase in $\chi^2/\text{d.o.f.}$
Although relatively smaller than the 0-parameter result, the
$\chi^2/\text{d.o.f.}$ of the 1-parameter NLO case (orange) also increases
beyond $k=0.1\ihMpc$, indicating a worsening of the fit from this point. Adding
a proxy $k^4$ counterterm to the NLO (purple), extends the range of
scales with good agreement for this NLO 2-parameter model to $k \simeq 0.15\ihMpc$. Nevertheless, it
is clear that the full NNLO 2-parameter model (red) is required to accurately
match Quijote data up to $k \simeq 0.18\ihMpc$. We note that when lowering
$\kmax$ below $0.1\ihMpc$, the fit becomes increasingly affected by sample
variance of the Quijote data, i.e.\ the available simulation data become
insufficient to reliably determine the EFT parameters in that case. We
therefore do not display lower $\kmax$ values in Fig.~\ref{fig:chi2}.

\subsection{Add $k^4$ counterterm to NNLO prediction}
\label{app:k4}

\begin{figure}
    \centering
    \includegraphics[width=0.8\textwidth]{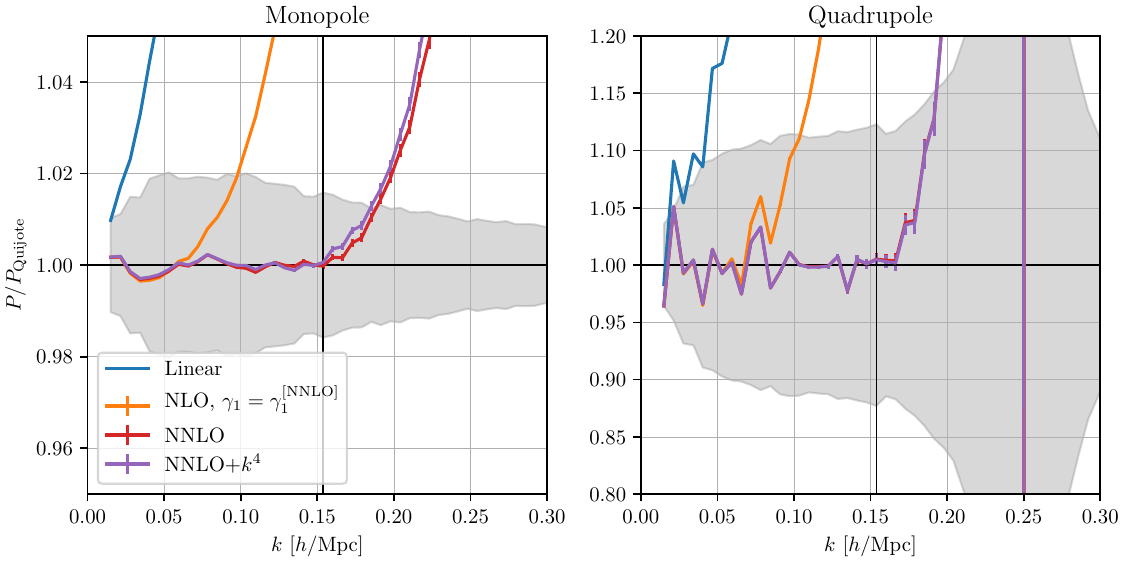}
        \caption{Impact of introducing a proxy $k^4 P_l^{\mathrm{lin}}(k)$
            counterterm at NNLO. The purple line shows the outcome of adding
            this counterterm, corresponding to a 3-parameter fit at NNLO. The
            0-parameter NLO (orange) and 2-parameter NNLO (red) are shown for
            comparison (with purple being almost on top of red lines, especially for the quadrupole). A pivot scale $\kmax = 0.15\ihMpc$ is used. All
            graphs are normalized to Quijote data, whose uncertainty is indicated by
            gray shading.}
    \label{fig:pk_nnlo_k4}
\end{figure}

We investigate whether adding a $k^4P_l^\text{lin}(k)$ counterterm to the NNLO result yields
any improvement. This 3-parameter NNLO model is plotted in
Fig.~\ref{fig:pk_nnlo_k4} (purple line), normalized to Quijote simulation data
at $z=0$. We find that adding the $k^4P_l^\text{lin}(k)$ term does not improve the fit compared
to the 2-parameter NNLO case. Note that the
leading contribution at small $k$ of the subtracted single-hard two-loop correction is in fact
proportional to $k^4 P_l^{\text{lin}}(k)$ (the double-hard limit $\propto k^2 P_l^{\text{lin}}(k)$ is
subtracted off). While this may be viewed to provide an explanation for the observed degeneracy of the single-hard
and $k^4$ counterterms, we note that the single-hard counterterm deviates from $k^4 P_l^{\text{lin}}(k)$ scaling already at $k\simeq 0.05\ihMpc$.
Nevertheless, within the region where it yields a sizable contribution we find that the $k$-dependence of the single-hard correction
is at least approximately close to a $k^4 P_l^{\text{lin}}(k)$ shape.
We also checked that adding the $k^4$ counterterm has no significant impact when increasing the pivot scale to $k_\text{max}=0.2\ihMpc$.

\subsection{Hexadecapole}
\label{app:hexadecapole}

\begin{figure}
    \centering
    \includegraphics[width=0.65\textwidth]{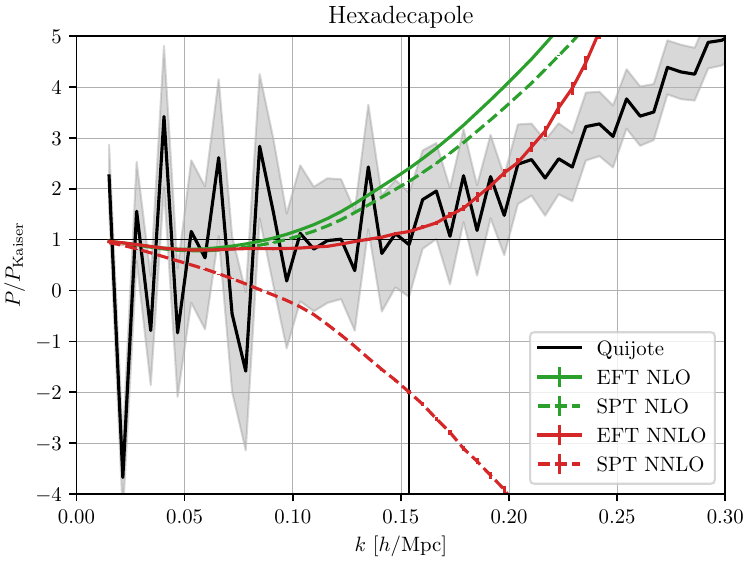}
        \caption{Perturbation theory predictions and Quijote simulation result
            for the power spectrum hexadecapole, normalized to the linear
            Kaiser prediction. Sample variance uncertainty from the $N$-body
            simulations is indicated with gray bands. The green solid lines
            correspond to a 0-parameter NLO prediction (as described in
            Sec.~\ref{sec:numerics}) and the red lines show our 2-parameter
            NNLO results. SPT results are displayed with dashed lines.}
    \label{fig:hexadecapole}
\end{figure}

Finally, in this section we apply our analysis to the hexadecapole in redshift
space. The results are displayed in Fig.~\ref{fig:hexadecapole}. Due to the
limited number of modes in the simulations on large scales, and their
distribution in $\mu$-space, the simulation estimate poorly recovers the
hexadecapole for $ k \lesssim 0.1\ihMpc$. Nevertheless, the EFT coefficients
at NNLO can be reasonably calibrated for $0.1 \ihMpc < k < 0.15 \ihMpc$. At
NLO, it is more challenging to reliably measure $\gamma_1$, therefore we show the
NLO result using $\gamma_1$ measured at NNLO. We can estimate that the NLO
result is in agreement with the simulations (within sample variance
uncertainty) up to $k \approx 0.14\ihMpc$ and the NNLO up to $k \approx
0.22\ihMpc$.

\section{Analytical results for the single- and double-hard limits}
\label{app:singledoublehard}

In this appendix we derive an analytical result for the single-hard  EFT correction $\bar P_{l,\text{sh}}^{\text{2-loop}}(k;\Lambda)$
defined in Eq.~\eqref{eq:P_l_2L_sh} that renormalizes
contributions to the two-loop power spectrum in redshift space for which one of the loop wavenumbers is large, see Sec.~\ref{sec:singlehardEFT}.
In Sec.~\ref{sec:Zninfty} we provide some intermediate results related to the single-hard limit for reference.

In addition, we provide an expression for the difference between the bare two-loop contribution ${P}_{l}^{\text{2-loop}}(k;\Lambda)$
and the quantity $\bar{P}_{l}^{\text{2-loop}}(k;\Lambda)$ used in our analysis, related to the double-hard limit, see Sec.~\ref{sec:doublehardEFT}.

\subsection{Single-hard EFT counterterm}\label{sec:singlehardEFT}

The integrand for the two-loop power spectrum in redshift space is given in Eq.~\eqref{eq:integrand}.
In the single-hard limit $q_1\to\infty$,
\bea
  p_{s,L=2}(\vk,\mu,\vq_1,\vq_2) &\to&  \frac{1}{q_1^2} \Big[  \left(30Z_5^\infty(\vk,\vq_2,-\vq_2)Z_1(\vk)+9Z_3^\infty(\vk)Z_3(\vk,\vq_2,-\vq_2)\right)P_0(k)\nn\\
  && {} + 12Z_4^\infty(\vk-\vq_2,\vq_2)Z_2(\vk-\vq_2,\vq_2)P_0(|\vk-\vq_2|)\Big] + {\cal O}(1/q_1^4)\,,
\eea
where we defined the asymptotic UV limit
\be
  Z_n^\infty(\vk_1,\dots,\vk_{n-2}) \equiv \int\frac{d\Omega_p}{4\pi} p^2 Z_n(\vk_1,\dots,\vk_{n-2},\vp,-\vp)|_{p\to \infty}\,,
\ee
which exists due to the scaling of the kernels $Z_n\propto p^{-2}+{\cal O}(p^{-4})$ imposed by mass and momentum conservation.
For $n=3$,
\bea
 Z_3^\infty(\vk) &=& -\frac{k^2}{1890}\left( 61 + 183 f \mu_k^2  + 3 f^2 \mu_k^2 (83 + 46 \mu_k^2) + 105 f^3 \mu_k^4\right)\,,
\eea
where $\mu_k=(\vk\cdot\hat z)/k$. Results for $n=4,5$ are given below. Apart from mass/momentum conservation constraints, we checked that our implementation of the $Z_n$
kernels satisfy Eq.~\eqref{eq:Zsoft}.

Using the definition of the single-hard EFT correction Eq.~\eqref{eq:P_l_2L_sh}, we find
\be
  P_{l,\text{sh}}^{\text{2-loop}}(k;\Lambda) =2\times 3 \times (A_l(k;\Lambda)+B_l(k;\Lambda))\,,
\ee
with
\bea
    A_l(k;\Lambda) &\equiv&
    \frac{2l + 1}{2}
    \int_{-1}^{1} \dd\mu_k \, \mathcal{P}_l(\mu_k)
    \int_{\vq_2}^\Lambda P_0(q_2)\left(30Z_5^\infty(\vk,\vq_2,-\vq_2)Z_1(\vk)+9Z_3^\infty(\vk)Z_3(\vk,\vq_2,-\vq_2)\right)P_0(k)\,,\nn\\
    B_l(k;\Lambda) &\equiv&
    \frac{2l + 1}{2}
    \int_{-1}^{1} \dd\mu_k \, \mathcal{P}_l(\mu_k)
    \int_{\vq_2}^\Lambda P_0(q_2)12Z_4^\infty(\vk-\vq_2,\vq_2)Z_2(\vk-\vq_2,\vq_2)P_0(|\vk-\vq_2|)\,.
\eea
Following~\cite{Baldauf:2015aha} we subtract the contribution to the single-hard limit that is already accounted for
by the leading $k^2P_0(k)$ EFT correction (see also Sec.~\ref{sec:doublehardEFT}). This amounts to using $\bar P_{l,\text{sh}}^{\text{2-loop}}(k;\Lambda) =2\times 3 \times (\bar A_l(k;\Lambda)+B_l(k;\Lambda))$
where $\bar A_l(k;\Lambda)$ differs from $A_l(k;\Lambda)$ by the replacement of $Z_5^\infty$ by
\be\label{eq:Z5inftybar}
  \bar Z_5^\infty(\vk,\vq_2,-\vq_2) \equiv Z_5^\infty(\vk,\vq_2,-\vq_2) - Z_5^\infty(\vk,\vq_2,-\vq_2)|_{q_2\to\infty}\,.
\ee
In order to perform the integration over $\mu_k$ we write $\mu_k=\hat z\cdot\vk/k$, where $\hat z$ is a generic unit vector in the direction
of the line-of-sight (not necessarily related to the third coordinate direction). Similarly, $\mu_{q_2}=\hat z\cdot\vq_2/q_2$. Due to rotational invariance of the (multipole-projected) power spectrum we can trade the integration over $\mu_k$ by an angular average
over $\hat z$, $\frac12\int d\mu_k\mapsto \int\frac{d\Omega_{\hat z}}{4\pi}$. This allows us to interchange the integration over $\vq_2$ with the average over the $\hat z$ direction.
After performing the latter, the integrand only depends on the magnitudes $k$ and $q_2$ as well as $c=\vk\cdot\vq_2/(kq_2)$. The integration over $c$ can then also be performed
for $\bar A_l$. Using the results from Sec.~\ref{sec:Zninfty} we find
\bea
    \bar A_l(k;\Lambda) &=&
    4\pi k^4 P_0(k)\int_0^\Lambda \dd q_2\, P_0(q_2) \bar a_l(k/q_2)\,,\nn\\
    B_l(k;\Lambda) &=&
    2\times 4\pi k^4 \int_0^\Lambda \dd q_2\, \frac12\int_{-1}^{\text{min}(1,k/(2q_2))} \dd c\, P_0(q_2)P_0(\sqrt{k^2+q_2^2-2kq_2c}) b_l(k/q_2,c)\,,
\eea
where we used the symmetry under $\vq_2\to \vk-\vq_2$ to limit the integration domain to $|\vk-\vq_2|>q_2$ for $B_l$, leading to
the additional factor of two and the upper limit for $c$. We find
\bea
  \bar a_l(x) &=& \frac{1}{444972528000 x^6}f_l(x)+\frac{(x^2-1)^3}{118659340800 x^7} \log\left(\frac{x+1}{x-1}\right)^2 g_l(x)\,,\nn\\
  b_l(x,c) &=& -\frac{1}{12978365400 (1 - 2 c x + x^2)^2}h_l(c,x)\,,
\eea
where
\bea
  f_0(x) &=&
  -122574375 + 1934903250 x^2 - 4063378900 x^4 + 6891099950 x^6  - 494386725 x^8 \nn\\
 && {} +  f (-245148750 + 3869806500 x^2 - 8126757800 x^4 + 13782199900 x^6 -     988773450 x^8)\nn\\
 && {} +  f^2 (-202153875 + 7398335770 x^2 - 16721366060 x^4 +     29724212070 x^6 - 2014729185 x^8) \nn\\
 && {}  +  f^3 (-35238060 + 8064808440 x^2 - 18792501312 x^4 +     37724885640 x^6 - 2540596500 x^8)\nn\\
 && {}  +  f^4 (36636600 + 3304341040 x^2 - 6875228360 x^4 + 30783953200 x^6 -     2116274160 x^8)\nn\\
 && {}  +  f^5 (378378000 x^2 - 504504000 x^4 + 11471860400 x^6 - 504504000 x^8)
 +  1373372000 f^6 x^6 \,,\nn\\
 g_0(x) &=& -8171625 + 107202550 x^2 + 32959115 x^4
  +  f (-16343250 + 214405100 x^2 + 65918230 x^4)\nn\\
 && {}   +  f^2 (-13476925 + 457283918 x^2 + 134315279 x^4)
  +  f^3 (-2349204 + 531389352 x^2 + 169373100 x^4)\nn\\
 && {}   +  f^4 (2442440 + 226802576 x^2 + 141084944 x^4)
  +  f^5 (25225200 x^2 + 33633600 x^4) \,,\nn\\
 h_0(x,c) &=& 13 \times \big[
   21635040 c^2 - 50500800 c^4 - 15798720 c x + 57731520 c^3 x +
   50500800 c^5 x - 58822470 c^2 x^2 \nn\\
 && {} \quad - 17613960 c^4 x^2 +
   29417640 c x^3 - 46938570 c^3 x^3 + 24166065 c^2 x^4 +
   315 (-6176 + 25933 x^2) \nn\\
 && {}
   +    f (43270080 c^2 - 101001600 c^4 - 31597440 c x + 115463040 c^3 x +
      101001600 c^5 x - 117644940 c^2 x^2 \nn\\
 && {} \quad - 35227920 c^4 x^2 +
      58835280 c x^3 - 93877140 c^3 x^3 + 48332130 c^2 x^4 +
      630 (-6176 + 25933 x^2))\nn\\
 && {}
   +    f^2 (-44344832 c^2 - 93438912 c^4 - 7964992 c x +
      275567488 c^3 x + 93438912 c^5 x - 373557016 c^2 x^2 \nn\\
 && {} \quad +
      70702856 c^4 x^2 + 156483950 c x^3 - 239765624 c^3 x^3 +
      100203334 c^2 x^4 \nn\\
 && {} \quad + 7 (2490944 + 6232451 x^2 + 230153 x^4)) \nn\\
 && {}
   +    f^3 (-88197888 c^2 - 36840960 c^4 + 17642496 c x +
      250077696 c^3 x + 36840960 c^5 x - 403308012 c^2 x^2 \nn\\
 && {} \quad +
      179937000 c^4 x^2 + 172087356 c x^3 - 321486564 c^3 x^3 +
      122697960 c^2 x^4 \nn\\
 && {} \quad + 12 (1959872 + 3689138 x^2 + 230153 x^4))\nn\\
 && {}
   +    f^4 (-23255232 c^2 - 3360896 c^4 + 36915648 c x + 53232256 c^3 x +
      3360896 c^5 x - 129287774 c^2 x^2 \nn\\
 && {} \quad + 121484440 c^4 x^2 +
      64401260 c x^3 - 205997330 c^3 x^3 + 81545849 c^2 x^4 \nn\\
 && {} \quad +
      77 (-59136 + 8797 x^2 + 62818 x^4))\nn\\
 && {}
   +    f^5 (5103840 c^2 x^2 + 26493320 c^4 x^2 - 1502340 c x^3 -
      58090480 c^3 x^3 + 27018110 c^2 x^4 \nn\\
 && {} \quad +
      51450 x^2 (-35 + 54 x^2))\nn\\
 && {}
   +    f^6 (2160900 c^2 x^2 + 1440600 c^4 x^2 - 2521050 c x^3 -
      5042100 c^3 x^3 + 3361400 c^2 x^4 \nn\\
 && {} \quad + 60025 x^2 (3 + 7 x^2))
   \big]
   \,,
\eea
\bea
 f_2(x) &=& 10\times\big[
    f (-49029750 + 773961300 x^2 - 1625351560 x^4 + 2756439980 x^6 -     197754690 x^8)\nn\\
 && {}   +  f^2 (-66284625 + 1936481900 x^2 - 4323380890 x^4 + 7488463140 x^6 -     515039685 x^8)\nn\\
 && {}   +  f^3 (-6344910 + 2439070140 x^2 - 5703792432 x^4 + 11373993540 x^6 -     769234050 x^8)\nn\\
 && {}   +  f^4 (17069325 + 1142872640 x^2 - 2450137690 x^4 + 9972981200 x^6 -     681352035 x^8)\nn\\
 && {}   +  f^5 (146191500 x^2 - 214987500 x^4 + 3959655700 x^6 - 174856500 x^8)\nn\\
 && {}   +  499408000 f^6 x^6 \big]\,,\nn\\
 g_2(x) &=& 10\times\big[
   f (-3268650 + 42881020 x^2 + 13183646 x^4)
 +  f^2 (-4418975 + 117314860 x^2 + 34335979 x^4)\nn\\
 && {}  +  f^3 (-422994 + 161476692 x^2 + 51282270 x^4)
 +  f^4 (1137955 + 79226056 x^2 + 45423469 x^4)\nn\\
 && {}  + f^5 (9746100 x^2 + 11657100 x^4)\big]\,,\nn\\
 h_2(x,c) &=& 5\times\big[
      f (225004416 c^2 - 525208320 c^4 - 164306688 c x +
      600407808 c^3 x + 525208320 c^5 x \nn\\
 && {} \quad - 611753688 c^2 x^2 -
      183185184 c^4 x^2 + 305943456 c x^3 - 488161128 c^3 x^3 +
      251327076 c^2 x^4 \nn\\
 && {} \quad + 3276 (-6176 + 25933 x^2))\nn\\
 && {}
  +    f^2 (-4713488 c^2 - 873754752 c^4 - 205533952 c x +
      1756936480 c^3 x + 873754752 c^5 x \nn\\
 && {} \quad - 2246440066 c^2 x^2 +
      215752394 c^4 x^2 + 996332246 c x^3 - 1554059234 c^3 x^3 +
      680064385 c^2 x^4 \nn\\
 && {} \quad + 13 (5390960 + 22198982 x^2 + 230153 x^4))\nn\\
 && {}
  +    f^3 (-504592608 c^2 - 481241280 c^4 + 15377856 c x +
      1971667776 c^3 x + 481241280 c^5 x \nn\\
 && {} \quad - 3104672272 c^2 x^2 +
      1273977640 c^4 x^2 + 1371167616 c x^3 - 2555882784 c^3 x^3 +
      991770780 c^2 x^4 \nn\\
 && {} \quad + 52 (3135992 + 7041278 x^2 + 230153 x^4))\nn\\
 && {}
  +    f^4 (-196948752 c^2 - 67523456 c^4 + 248696448 c x +
      528944416 c^3 x + 67523456 c^5 x \nn\\
 && {} \quad - 1267193564 c^2 x^2 +
      1145203150 c^4 x^2 + 661661000 c x^3 - 1918651280 c^3 x^3 +
      745214834 c^2 x^4 \nn\\
 && {} \quad + 182 (-94776 + 253747 x^2 + 132643 x^4))\nn\\
 && {}
  +    f^5 (-15887760 c^2 x^2 + 311402840 c^4 x^2 + 38073000 c x^3 -
      606917920 c^3 x^3 + 267402800 c^2 x^4 \nn\\
 && {} \quad +
      20580 x^2 (-539 + 827 x^2))\nn\\
 && {}
  + f^6 (16206750 c^2 x^2 + 20888700 c^4 x^2 - 17647350 c x^3 -
      57984150 c^3 x^3 + 34874525 c^2 x^4 \nn\\
 && {} \quad + 60025 x^2 (12 + 49 x^2))
      \big]\,.
\eea
Note that all $a_l$ and $b_l$ approach a finite limit for both $x\to 0$ (i.e. $q_2\gg k$) and $x\to\infty$  (i.e. $q_2\ll k$).
In the latter limit, we checked that the constant terms cancel in the sum $a_l(x)+\int_{-1}^1 \dd c\, b_l(x,c)\propto 1/x$,
as required by Galilean invariance, and similarly to the usual analogous cancellation at one-loop order.

For the various angular averages we used
\be
  \int \frac{\dd\Omega_q}{4\pi} (\hat v_1\cdot\hat q)^{n_1}\cdots(\hat v_N\cdot\hat q)^{n_N} = \left(\frac{\partial}{\partial\lambda_1}\right)^{n_1}\cdots\left(\frac{\partial}{\partial\lambda_N}\right)^{n_N}\frac{\sinh|\sum_i\lambda_i\hat v_i|}{|\sum_i\lambda_i\hat v_i|}\Big|_{\lambda_i=0}\,,
\ee
which can be readily shown by computing $\int \frac{\dd\Omega_q}{4\pi} e^{\sum_i\lambda_i\hat v_i\cdot\hat q}$. Here $\hat q=\vq/|\vq|$.

\subsection{Single-hard limit of $Z_4$ and $Z_5$}\label{sec:Zninfty}

\bea
 Z_4^\infty(\vk-\vq,\vq) &=& -\frac{k^2}{4074840 (1 - 2 c x + x^2) }\Big[
   -  6176 + 48096 c^2 - 48096 c^3 x + 25933 x^2 - 16892 c^2 x^2
     \nn\\ && {} \quad
     + c x (-35744 + 32879 x^2)
  \nn\\ && {}
   +  f \mu_k (11 \mu_q (16032 c - 8016 x - 16032 c^2 x + 2038 c x^2 +        2989 x^3) +     \mu_k (-16032 c^3 x
     \nn\\ && {} \quad
     + c^2 (16032 - 89986 x^2) +        4 (-6176 + 25933 x^2) + c x (-54800 + 98637 x^2)))
  \nn\\ && {}
   +  33 f^2 \mu_k^2 (4752 + 7443 x^2 + 4190 \mu_k^2 x^2 -     2 c^2 (3379 + 539 \mu_k^2) x^2 +     \mu_q^2 (5456 - 5456 c x + 2377 x^2)
     \nn\\ && {} \quad
      +     \mu_k \mu_q x (-5456 - 5276 c x + 2989 x^2) +     c x (-9504 + 49 (83 + 46 \mu_k^2) x^2))
  \nn\\ && {}
   +  1617 f^3 \mu_k^3 x^2 (46 \mu_k^2 \mu_q x + 83 \mu_q (-2 c + x) +     \mu_k (78 - 30 c^2 - 46 \mu_q^2 + 35 c x))
  \nn\\ && {}
   +  56595 f^4 \mu_k^4 \mu_q x^2 (-\mu_q + \mu_k x)
 \Big]\,,
\eea
\bea
 Z_5^\infty(\vk,\vq,-\vq) &=& -\frac{k^2}{264864600 (1 - 2 c x + x^2)(1 + 2 c x + x^2) }\Big[
    1832216 c^2 - 173536 c^4 + 1763891 c^2 x^2
     \nn\\ && {} \quad
      - 6602262 c^4 x^2 +  382844 c^6 x^2 + 861688 c^2 x^4 + 1462930 c^4 x^4
     \nn\\ && {} \quad
      - 427427 c^2 x^6 +  4 (1 + x^2) (32614 + 79843 x^2)
  \nn\\ && {}
    +  f \mu_k (26 c \mu_q (299792 + 496133 x^2 + 69608 c^4 x^2 + 223706 x^4 -
       32879 x^6
     \nn\\ && {} \quad
         + 4 c^2 (-7888 - 283775 x^2+ 27573 x^4)) +
       \mu_k (104412 c^6 x^2 + 20 (32614
     \nn\\ && {} \quad
       + 112457 x^2 + 79843 x^4) +
       c^4 (-47328 - 3498710 x^2 + 4447058 x^4)
     \nn\\ && {} \quad
       -        c^2 (-1366488 + 4080003 x^2 + 1507916 x^4 + 1282281 x^6)))
  \nn\\ && {}
    +  13 f^2 \mu_k^2 (-197960 c^3 \mu_k \mu_q x^2 + 774232 c^3 \mu_k \mu_q x^4 -
      2 c \mu_k \mu_q (-76192 + 152041 x^2
     \nn\\ && {} \quad
      + 113650 x^4 + 98637 x^6) +
      4 c^4 x^2 (-10811 + 45866 \mu_q^2 + 127953 x^2 +
       98 \mu_k^2 (43 + 568 x^2))
     \nn\\ && {} \quad
       -       c^2 (-62072 + (2392381 + 1073122 \mu_k^2) x^2 +
       4 (67831 + 53144 \mu_k^2) x^4 + 1617 (83 + 46 \mu_k^2) x^6
     \nn\\ && {} \quad
       +        \mu_q^2 (75016 + 2733892 x^2 - 84136 x^4)) - (-1 -
       x^2) (521816 + 611822 x^2
     \nn\\ && {} \quad
       + 4 \mu_k^2 (52700 + 87371 x^2) -
       7 \mu_q^2 (-97432 - 88787 x^2 + 4697 x^4)))
  \nn\\ && {}
    +  429 f^3 \mu_k^3 (56 c^3 \mu_q (249 + 61 \mu_q^2) x^2 +     392 c^3 (83 + 35 \mu_k^2) \mu_q x^4 + 980 c^4 \mu_k x^2 (1 + 5 x^2)
     \nn\\ && {} \quad
      -     c^2 \mu_k x^2 (28185 + 7952 x^2 + 1715 x^4 -        4 \mu_q^2 (-5344 + 3101 x^2)) -     14 c \mu_q x^2 (244 \mu_q^2
     \nn\\ && {} \quad
       + 83 (19 + 14 x^2 + 7 x^4) +
       14 \mu_k^2 (23 + 24 x^2 + 23 x^4)) +
      \mu_k (-1 - x^2) (-2 (2376 + 5617 x^2)
     \nn\\ && {} \quad
      +        \mu_q^2 (-5456 - 2019 x^2 + 2989 x^4)))
  \nn\\ && {}
    +  21021 f^4 \mu_k^4 \mu_q x^2 (160 c^3 \mu_k x^2 +     4 c^2 \mu_q (15 + (98 + 46 \mu_k^2) x^2) -     10 c \mu_k (7 + 2 x^2 + 7 x^4)
     \nn\\ && {} \quad
     -     \mu_q (1 + x^2) (143 + 83 x^2 + 46 \mu_k^2 (1 + x^2)))
  \nn\\ && {}
    -  735735 f^5 \mu_k^6 \mu_q^2 x^2 (1 + (2 - 4 c^2) x^2 + x^4)
 \Big]\,.
\eea
We used the abbreviations $x=k/q$ and $c=\vk\cdot\vq/(kq)$.
We checked that our results for $Z_n^\infty$ for $n=3,4,5$ agree with those for the corresponding density kernels $F_n^\infty$ (see e.g.~\cite{Baldauf:2015aha,Baldauf:2021zlt}) in the formal limit
when setting the growth rate $f$ to zero. We also used Eq.~\eqref{eq:Zsoft} for further cross-checks in the limit $q\ll k$ ($x\to\infty$).

\subsection{Double-hard limit}\label{sec:doublehardEFT}

Contributions to the two-loop power spectrum for which both wavenumbers are large, $q_1,q_2\gg k$, are degenerate with the EFT counterterm $\gamma_1$ proportional to
$k^2P_0(k)$ that was already introduced to renormalize the one-loop result. We therefore choose the option to subtract these contributions in our analysis. Note that
this is not strictly necessary, and would merely lead to different values of $\gamma_1$. The subtracted two-loop contribution is given by
\be
  \bar{P}_{l}^{\text{2-loop}}(k;\Lambda) \equiv {P}_{l}^{\text{2-loop}}(k;\Lambda) - {P}_{l,\text{dh}}^{\text{2-loop}}(k;\Lambda)\,.
\ee
The double-hard limit is given by
\be
  {P}_{l,\text{dh}}^{\text{2-loop}}(k;\Lambda) = \frac{2l + 1}{2}
    \int_{-1}^{1} \dd\mu_k \, \mathcal{P}_l(\mu_k)
    \int_{\vq_1}^\Lambda\int_{\vq_2}^\Lambda P_0(q_1)P_0(q_2)30Z_5(\vk,\vq_1,-\vq_1,\vq_2,-\vq_2)|_{q_i\to\infty}Z_1(\vk)P_0(k)\,,
\ee
where the $Z_5$ kernel is evaluated in the limit where both loop wavenumbers becomes large, with fixed ratio $q_1/q_2$.
It can be written as
\be\label{eq:Pdoublehard}
  {P}_{l,\text{dh}}^{\text{2-loop}}(k;\Lambda) =
    30k^2P_0(k)\int_{\vq_1}^\Lambda\int_{\vq_2}^\Lambda P_0(q_1)P_0(q_2)\frac{1}{q_1q_2}Z^{15}_{l,\text{dh}}(q_1/q_2)\,.
\ee
Using techniques similar as above we find
\be
  Z^{15}_{l,\text{dh}}(x) = \frac{1+x^2}{238378140000 x^5}p_l(x)+\frac{(x^2-1)^4}{63567504000 x^6} \log\left(\frac{x+1}{x-1}\right)^2 q_l(x)\,,
\ee
where here $x=q_1/q_2$ with
\bea
p_0(x) &=&
    1155 (5760 + 13605 x^2 - 128258 x^4 + 13605 x^6 + 5760 x^8)
  \nn\\ && {}
  +  2310 f (5760 + 13605 x^2 - 128258 x^4 + 13605 x^6 + 5760 x^8)
  \nn\\ && {}
  +  7 f^2 (1193760 + 31353255 x^2 - 159738358 x^4 + 31353255 x^6 +     1193760 x^8)
  \nn\\ && {}
  +  234 f^3 (6240 + 660295 x^2 - 3151542 x^4 + 660295 x^6 + 6240 x^8)
  \nn\\ && {}
  +  7078500 f^4 x^2 (3 - 14 x^2 + 3 x^4)
  \,,\nn\\
q_0(x) &=&
  - 1155 (384 + 2699 x^2 + 384 x^4)
  -  2310 f (384 + 2699 x^2 + 384 x^4)
  \nn\\ && {}
  -  7 f^2 (79584 + 2461609 x^2 + 79584 x^4)
  -  234 f^3 (416 + 45961 x^2 + 416 x^4)
  -  1415700 f^4 x^2
  \,,\nn\\
p_2(x) &=&
  10  \big[
        462 f (5760 + 13605 x^2 - 128258 x^4 + 13605 x^6 + 5760 x^8)
  \nn\\ && {}
    +    11 f^2 (233640 + 4284945 x^2 - 22623562 x^4 + 4284945 x^6 +       233640 x^8)
  \nn\\ && {}
    +    13 f^3 (42120 + 3453585 x^2 - 16590346 x^4 + 3453585 x^6 +       42120 x^8)
  \nn\\ && {}
    +   2359500 f^4 x^2 (3 - 14 x^2 + 3 x^4)
  \big]
  \,,\nn\\
q_2(x) &=&
  -10  \big[
   462 f (384 + 2699 x^2 + 384 x^4)
  +  11 f^2 (15576 + 358351 x^2 + 15576 x^4)
  \nn\\ && {}
  +  13 f^3 (2808 + 243343 x^2 + 2808 x^4)
  + 471900 f^4 x^2
  \big]
  \,.
\eea
Note that $Z^{15}_{l,\text{dh}}(x)=Z^{15}_{l,\text{dh}}(1/x)$.
For $f=0$ this result agrees with the double-hard limit of the real space power spectrum (see e.g.~\cite{Baldauf:2015aha,Baldauf:2021zlt}).
Furthermore, we checked a correspondence of the double-hard limit and the ``hard limit of the single-hard limit''.
More precisely, we checked that for the single-hard contribution, the difference between $P_{l,\text{sh}}^{\text{2-loop}}(k;\Lambda)$ and
$\bar P_{l,\text{sh}}^{\text{2-loop}}(k;\Lambda)$ (see subtraction term in Eq.~\eqref{eq:Z5inftybar}) can be related to Eq.~\eqref{eq:Pdoublehard}
evaluated with $Z^{15}_{l,\text{dh}}(x)|_{x\to \infty}$. Note that the leading term in  $Z^{15}_{l,\text{dh}}(x)$ is scaling as $\propto 1/x$ for large $x$,
such that the integrand in Eq.~\eqref{eq:Pdoublehard} scales as $1/(q_1q_2x)=1/q_1^2$.

% \bibliographystyle{JHEP}
% \bibliography{main}

\providecommand{\href}[2]{#2}\begingroup\raggedright\endgroup

\end{document}